\documentclass[
footinbib,
a4paper,
aps,
prx,
superscriptaddress,
reprint,
twocolumn,
preprintnumbers,
nobalancelastpage,
floatfix,
10pt]{revtex4-2}

\usepackage{dutchcal}

\usepackage{xcolor}
\usepackage{graphicx}
 	\graphicspath{ {Figures/} }
\usepackage{dcolumn}
\usepackage{tikz}
\usepackage{subfigure}
\usepackage{float}
\definecolor{cmblue}{rgb}{0.12156862745098039, 0.4666666666666667, 0.7058823529411765}
\definecolor{mygrey}{gray}{0.35}
\definecolor{myblue}{rgb}{0.2,0.2,0.8}
\definecolor{mygreen}{rgb}{0.2,0.8,0.5}
\definecolor{myzard}{cmyk}{0,0,0.05,0}
\definecolor{mywhite}{rgb}{1,1,1}
\definecolor{myred}{rgb}{1,0.,0.3}

\usepackage[utf8]{inputenc}
\usepackage{dsfont}
\usepackage[normalem]{ulem}
\usepackage{bbold}
\usepackage{soul}
\usepackage{slashed}

\usepackage{amsmath, amsthm, amsfonts, amssymb,amstext}
\usepackage{mathptmx}
\usepackage{bm}
\usepackage{xfrac}
\usepackage{mathrsfs}
\usepackage{latexsym}
\usepackage{array}

\usepackage{makecell}

 \def\ee{\mathord{\mathrm e}}
 
 \def\ii{\mathord{\mathrm i}}

\def\beq{\begin{equation}}
\def\eeq{\end{equation}}
\def\barray{\begin{eqnarray}}
\def\earray{\end{eqnarray}}

\usepackage{physics}
\usepackage{braket}
\usepackage{tikz}

\usepackage[
colorlinks=true,
citecolor=myblue,
linkcolor=myred]{hyperref}

\usepackage{cleveref} 
\usepackage{verbatim}

\usetikzlibrary{backgrounds,fit,decorations.pathreplacing,calc}
\usetikzlibrary{quantikz2}
\renewcommand{\ii}{{\mathrm i}}
\renewcommand{\ee}{{\mathrm e}}

\renewcommand{\eqref}[1]{\textup{Eq.~(\ref{#1})}}

\begin{document}

\author{Pablo Viñas}
\affiliation{Instituto de Física Teórica (UAM-CSIC), Universidad Autónoma de Madrid, Cantoblanco, 28094 Madrid, Spain}
\email{pablo.vinnas@estudiante.uam.es}

\author{Alan Kahan}
\affiliation{Instituto de Física Teórica (UAM-CSIC), Universidad Autónoma de Madrid, Cantoblanco, 28094 Madrid, Spain}

\author{Tommaso Faorlin}
\affiliation{Universität Innsbruck, Institut für Experimentalphysik, Technikerstraße 25/4, 6020 Innsbruck, Austria}

\author{Lorenz Panzl}%
\affiliation{Universität Innsbruck, Institut für Experimentalphysik, Technikerstraße 25/4, 6020 Innsbruck, Austria}

\author{Phoebe Grosser}%
\affiliation{Universität Innsbruck, Institut für Experimentalphysik, Technikerstraße 25/4, 6020 Innsbruck, Austria}

\author{Thomas Monz}
\affiliation{Universität Innsbruck, Institut für Experimentalphysik, Technikerstraße 25/4, 6020 Innsbruck, Austria}
\affiliation{Alpine Quantum Technologies, Technikerstraße 17, 6020 Innsbruck, Austria}

\author{Alejandro Bermudez}%
\affiliation{Instituto de Física Teórica (UAM-CSIC), Universidad Autónoma de Madrid, Cantoblanco, 28094 Madrid, Spain}

\title{Structured light for addressable two-qubit gates  in medium-sized ion crystals:\\ modeling motional and dephasing errors}

\begin{abstract}
Geometric phase gates with addressed laser radiation are widely used in trapped-ion quantum computers to drive 
high-fidelity entangling gates on an arbitrary qubit pair of a larger ion chain. Such programmable connectivity exploits the collective phonons to mediate the desired interaction, while simultaneously achieving robustness to their thermal fluctuations by a careful calibration of the gate time with respect to the laser and vibrational-mode frequencies. Crosstalk errors can arise when the spectator qubits, which should ideally remain idle during the gate, are partially driven by the laser fields. The minimal crosstalk, which is set by the laser beam size on the spectator ions, can be achieved when the addressing beams propagate perpendicular to the ion chain. However, this results in exciting the radial phonons, and can complicate achieving high-fidelities as the ion chain grows due to the so-called spectral mode crowding. In this work, we present the  theoretical description of a scheme recently implemented by Faorlin et {\it al.}~\citenum{faorlin2026entanglingionsengineeredlight}. This scheme overcomes these complications by engineering addressable state-dependent forces along the ion-chain with laser beams that propagate perpendicular to it through the use of structured light gradients. In this way, one can excite the axial phonons and avoid the spectral mode crowding, while simultaneously minimizing the crosstalk.  
We present a detailed derivation of the entangling gate scheme, together with analytical closed-form expressions for the leading contribution to the gate error  caused by thermal motional fluctuations and dephasing noise. 
Our results confirm that such fluctuations account for up to 93$\%$ of the error budget under the operating conditions of Ref.~\citenum{faorlin2026entanglingionsengineeredlight}. We identify target parameters to reduce the motional error from the reported $10^{-3}$ to the $10^{-5}$ level.
\end{abstract}

\maketitle

\section{Introduction}
\label{sec. introduction}
Entangling gates in quantum processing units (QPUs) exhibit higher error rates than their single-qubit counterparts \cite{Baldwin2021, Ballance2016} which, in turn, limits the depth of quantum circuits of near-term applications. With current error rates slightly below the fault-tolerant threshold of quantum error correction (QEC),  the overhead in the number of physical qubits required to demonstrate a logical-to-physical advantage  is still daunting for most current setups~\cite{Campbell_2017}.
Therefore, improving the fidelity of the entangling gates is regarded as one of the most effective ways to increase circuit depth or reduce QEC overheads and, thus, to enable future quantum applications demonstrating practical quantum advantage. Trapped-ion QPUs offer the highest fidelity entangling gates when operating on small ion crystals, and a clear goal is to maintain these high fidelities while the size of the quantum register increases. 
In practice, the fidelity of two-qubit entangling gates degrades with the size of the ion crystal. As a means of addressing this limitation, quantum charge-coupled devices (QCCD) architectures~\cite{kielpinski2002architecture} were conceived to allow for  splitting,
 shuttling and merging of ion crystals, bringing arbitrary pairs into dedicated interaction zones of the trap where high-fidelity two-qubit operations can be driven~\cite{Kielpinski, Home_2009, PhysRevX.13.041052, Kaushal2020, Hilder2022}. 
However, these shuttling-type operations tend to consume a considerable portion of the compute time required to implement a specific quantum circuit. 
Modular QCCD architectures can benefit from high-fidelity gates operated in medium-sized ion crystals as building blocks, as this partially reduces the shuttling overhead incurred when entangling gates are only driven in two-ion crystals. In this work, we derive a theoretical error budget of an entangling gate, introduced and experimentally realised in Ref.~\citenum{faorlin2026entanglingionsengineeredlight}, where error rates at the $10^{-3}$ level were demonstrated with up to $N=12$ ions by exploiting structured light beams. As we show theoretically, considering the same microscopic control parameters, the contribution of thermal motional fluctuations and dephasing noise to the gate error remains at this level for chains with up to $N=20$ ions confined within a single potential well. Importantly, this scalability is achieved while preserving optimal single-ion addressing.

We focus on geometric phase gates, which  are usually realized via state-dependent forces (SDFs) that couple the qubits to the vibrational modes of the ion chain \cite{Leibfried2003, Sorensen1999}. By carefully adjusting this SDF, the phonon modes that mediate the qubit-qubit interactions are driven along closed phase-space trajectories that depend on the collective state of the spins through the enclosed geometric area, such that entanglement between electronic states can be generated with high fidelity. Achieving this requires minimizing the effect that the thermal fluctuations on the phonon number can have on the entangling process.
When using a single SDF with fixed intensity/phase, either a single or a pair of vibrational modes act as the quantum data bus carrying the effective qubit-qubit interactions~\cite{_A_URA_2002}, while the remaining spectator modes give rise to off-resonant couplings that make the dynamics  deviate from the ideal unitary gate~\cite{S_rensen_2000}.  
As the number of vibrational modes grows with the chain size, the number of off-resonant couplings to such modes increases, leading to a degradation of gate performance. This effect is specifically prominent when the SDF acts on the radial part of the motional spectrum since, contrary to axial modes, radial modes become closely spaced in frequency. As a consequence, the strength of the corresponding parasitic couplings and the residual qubit-phonon entanglement become larger.

A common optical setup providing the SDF is based on illuminating the desired ions with lasers in a traveling or standing wave configuration~\cite{_A_URA_2002}.
The recoil momentum imparted on the ions determines which phonons, either axial or radial, couple to the qubits and mediate the entangling gate~\cite{Leibfried2003, Sorensen1999, PhysRevA.109.052417}.
For a global-beam scheme, selecting the desired qubit pair for the entangling gate requires additional experimental resources for laser-addressed shelving~\cite{Nigg2014} or selective recoupling~\cite{Leung01072002}, which can also  to errors from the imperfect spectroscopic decoupling-recoupling pulses or the nuclear magnetic resonance (NMR)-type refocusing ones. 
As an alternative, one can use addressed-beam schemes that directly focus the laser light responsible for the two-qubit gates on the desired ion pair. Such schemes minimize the crosstalk to the remaining neighboring qubits in the chain by directing the light perpendicularly to the ion chain. In this situation, however, the standard schemes create an SDF that only excites the radial vibrational modes, and
one needs to deal with the spectral crowding to achieve high fidelities. We note that various schemes pulsing the SDF or shaping its envelope and phase have been developed for this goal, leading to increased control complexity and calibration requirements~\cite{PhysRevApplied.13.024022, PhysRevLett.112.190502, PhysRevLett.114.120502, Steane_2014, PhysRevLett.120.020501, PhysRevA.98.032318, PhysRevA.100.022332, Sch_fer_2018, 10.1038/s41586-019-1428-4}. 

Recently, there has been an increased interest in using structured light to mediate such entangling interactions and, more specifically, in exploring its application in quantum computing~\cite{West2021, Mazzanti2023, Cui2025, Mai2026}. Using structured light allows single-ion addressing~\cite{faorlin2026entanglingionsengineeredlight, Mai2026} while creating an SDF that couples the qubits to the axial modes~\cite{Stopp2022} allows for the realization of entangling gates that  are less susceptible to spectral crowding for certain laser frequencies~\cite{Zhu2006, Mai2026}.
In Ref.~\citenum{faorlin2026entanglingionsengineeredlight}, we presented a scheme for the generation of  structured-light SDF and demonstrated high-fidelity geometric phase gates for  crystals of up to $N=12$ ions.
There, the qubit-phonon coupling was obtained by interfering a pair of transversal electromagnetic (TEM) modes with different spatial profiles. This results in either an effective polarization or intensity gradient at the position of the addressed ions, which can be oriented parallel to the ion chain and transverse to the propagation direction of the light beams. 

In this article, we present a detailed theoretical account of this scheme, and perform a thorough analysis of the gate dynamics. We confirm that the proposed approach is less sensitive to the coupling to spectator modes than other addressed implementations of entangling gates. We also show that high fidelities are maintained when increasing the ion number up to $N=20$. We provide a detailed account of two fundamental sources of noise in the scheme: thermal motional fluctuations of various vibrational modes, and stochastic phase noise. In particular, we analyze the effects of higher-order terms in the Lamb–Dicke expansion, as well as  off-resonant couplings to spectator modes mediated by the polarization gradient and ion recoil. In addition, we also account for the  change in the transition frequency of the qubit due to magnetic field noise. Our analysis provides further insights into experimentally relevant gate properties, including the identification of operating parameters that optimize gate performance. Moreover, since the formalism is predominantly analytical, it allows for the exploration of gate configurations in medium-sized ion chains that are beyond the reach of direct numerical simulations.

This work is organized as follows. We start by reviewing the geometric phase gates in Sec.~\ref{sec. working_principle} to set the stage, define our notation and establish the theory behind the implementation of the transversely-addressed structured-light  gate. Then, in section~\ref{sec. error_analysis}, we develop a precise error analysis including the infidelity contributions derived from the approximations made to obtain the ideal gate description in the former section. This analysis is divided into two subsections: Subsec.~\ref{subsec. fidelity1}, which accounts for infidelity contributions associated to the active modes addressed to drive the gate, and Subsec.~\ref{subsec. fidelity2}, in which we extend the analysis to spectator modes and perform realistic fidelity estimations for the gate. The error analysis continues in Sec.~\ref{sec: dephasing}, in which the dephasing produced by magnetic noise fluctuations of the qubit frequency is included, accounting for its contribution to the gate  infidelity. Based on these results, we conclude that this approach enables high-fidelity entanglement in medium-sized trapped-ion chains, and discuss its implications for applications such as QEC in Sec.~\ref{sec. conclusions}.

\section{Geometric phase gates}
\label{sec. working_principle}

Geometric-phase gates generate entanglement by exploiting SDFs. The application of these forces to a pair of trapped ions induces state-dependent trajectories of the vibrational modes in phase space. Although the physical implementation is very different, both Mølmer-Sørensen (MS)~\cite{Sorensen1999} and light-shift (LS)~\cite{Leibfried2003} gates are examples of such geometric-phase gates, with their respective force being dependent on a different Pauli basis. In the interaction picture, and assuming that only a single vibrational mode acts as the quantum data bus in a two-ion chain, the Hamiltonian leading to the SDF is~\cite{Sorensen1999, milburn_warm_ions, PhysRevLett.91.157901, Leibfried2003, Lee_2005, Ballance2017}
\begin{equation}
\label{eq: sd_force}
H(t)=\ii\mathcal{F}a^{\dagger}\ee^{\ii\delta t}\Lambda+\mathrm{H.c.},
\end{equation}
and the free Hamiltonian is $H_0=\nu a^{\dagger}a+\frac{1}{2}\omega_0(Z_1+Z_2)$.  Here we denote the $j$-th spin Pauli-$z$ operator by $Z_j$. The detuning $\delta=\nu-\nu_d$ is defined by the difference between the normal mode frequency $\nu$ and the forcing frequency $\nu_d$, whereas $\mathcal{F}$ is proportional to the amplitude of the force and $a^{\dagger}$ the creation operator of phonons in the vibrational mode that mediates the entangling gate. Additionally, $\Lambda$ is a linear combination of Pauli operators, and depends on the nature of the experimental scheme. We use $\mathrm{H.c.}$ to denote the Hermitian conjugate. 

This Hamiltonian yields an evolution that is governed by the unitary~\cite{Lee_2005, Ballance2017, PhysRevA.109.052417}
\begin{equation}
\label{eq: sd_evolution}
U(t)=e^{-i H_0t}e^{(\alpha(t)a^{\dagger}-\mathrm{H.c.})\Lambda}e^{-i \Phi(t)\Lambda^2},
\end{equation}
with $\alpha(t)=\ii\frac{\mathcal{F}}{\delta}(1-\ee^{\ii\delta t})$ being the argument of the displacement operator for the vibrational mode, which describes the trajectory in phase space. We also introduced a spin-spin coupling strength leading to the pulse area
\begin{equation}
\Phi(t)=-\frac{\mathcal{F}^2}{\delta}\left(t-\frac{\sin\delta t}{\delta}\right).
\end{equation}

Notably, when the duration of the forcing is adjusted to $\mathrm{t}_\mathrm{gate}=2\pi N_{\mathrm{loops}}/\delta$ and the amplitude is $\mathcal{F}=\delta/(2\sqrt{N_{\mathrm{loops}}})$, the fully entangling operator $U(\mathrm{t}_\mathrm{gate})=e^{-\ii H_0\mathrm{t}_\mathrm{gate}}e^{\ii\frac{\pi}{2}\Lambda^2}$ is obtained. Here, $N_{\mathrm{loops}}$ denotes the number of loops that the phase-space trajectory of the vibrational modes undergoes during the gate duration. Specially relevant to our work are the gates in which the force is dependent on the computational basis, such that $\Lambda=\frac{1}{2}(Z_1+Z_2)$. The archetypal implementation of this gate corresponds to the aforementioned LS gate, which produces maximally entangled states in this basis.

\subsection{Structured light gradient for entangling gates}
\label{subsec. implementation}

\begin{figure*}[t!]
\centering
\includegraphics[width=\textwidth]{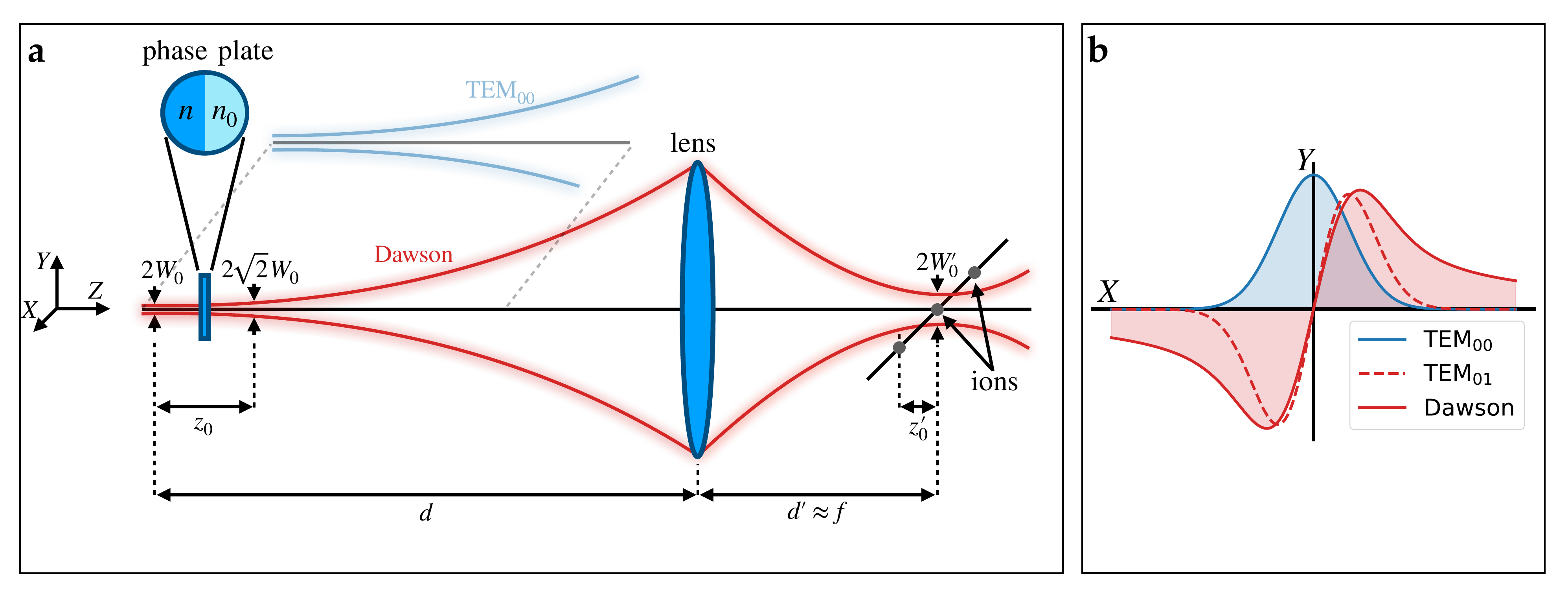}
\caption{\textbf{a) Scheme of the optical setup.} A gaussian beam with beam waist $W_0$ and Rayleigh range $z_0$ travels in the positive $z$ direction. An odd spatial profile is produced by passing the beam through a phase plate, an optical element with refractive index $n$ for $x<0$, and $n_0=1$ otherwise. The refractive index and the length of the plate are chosen such that the beam accumulates a phase difference of $\Delta\phi=\pi$ between the two halves. The beam travels a distance $d$ before reaching a lens with focal length $f$ and which focuses the beam at the ion's position, which is approximated by $d+f$ when the beam reaches the lens in a paraxial regime. Under the Fresnel approximation, the beam is described by the Fourier transform property of a lens at the focal plane~\cite{doi:https://doi.org/10.1002/0471213748.ch4, Goodman1969IntroductionTF}, and the complex amplitude is proportional to a Dawson function along the $x$ direction. This beam is represented in red. Similarly, the gaussian ($\mathrm{TEM}_{00}$) beam is represented in blue. The gaussian beam with symmetric complex amplitude does not pass through the phase plate. \textbf{b) TEM modes comparison.} The gaussian mode ($\mathrm{TEM}_{00}$) and anti-symmetric mode complex amplitudes are plotted in solid blue and solid red lines, respectively. The $\mathrm{TEM}_{01}$ mode is plotted with a dotted red line for comparison.}
\label{fig: optical_scheme}
\end{figure*}

In this subsection, we describe the basic principle underlying the implementation of the transversely-addressed light gradient gate, and the approximations which lead to the ideal time-evolution of \eqref{eq: sd_force}.
Usually, an effective entangling interaction between qubit states is mediated by the vibrational modes of the ion chain,
but the approach to engineer the dynamics described in above is not unique. We focus here on an alternative route: generating a position-dependent Rabi frequency through the optical generation and combination of structured transversal light modes that resemble the Laguerre-Gauss eigenmodes of an optical resonator.
In a recently realized experiment~\cite{faorlin2026entanglingionsengineeredlight}, a combination of traveling spatially symmetric and anti-symmetric beams with different frequencies (see Fig.~\ref{fig: optical_scheme}), generated using passive beam-shaping optics, produces an oscillating intensity or polarization gradient at the ion positions. The gradient is oriented along the ion-chain axis, while the beams propagate transversely to the chain, thereby optimizing the selectivity of the addressing. The modulation gives rise to the oscillating force that drives the entangling gate.

For a detailed description of the scheme, let us first consider a single ion interacting with a laser field close to resonance with one of its internal electronic transitions, described by the following interaction Hamiltonian
\begin{equation}
    H= \frac{\hbar}{2}\Omega(x)e^{-i \Delta t}\sigma_+  +  \mathrm{H.c.}.
\end{equation}

Here $\Omega(x)$ is the single photon Rabi frequency, $\sigma_+=|1\rangle\langle 0 |$ is the spin raising operator, and $\Delta$ is the detuning between the laser frequency and the transition frequency. The ion undergoes small fluctuations $\delta x$ from its classical equilibrium position $x^{\textrm{eq}}$, which, after cooling, and considering a single direction of motion, are well described by the quantum mechanical operator $\delta \hat{x}= \frac{1}{\sqrt{2m\nu_s}} (a_s^\dagger +a_s)$~\cite{_A_URA_2002}, where $m$ is the ion mass. We denote the ion trap's secular motion frequency by $\nu_s$ and the creation and annihilation operators are denoted by $a^\dagger_s$ and $a_s$, respectively. The subscript $s$ distinguishes single-ion motional quantities from the corresponding ones associated with the collective motional mode introduced in~\eqref{eq: sd_force}. After a Taylor expansion of the position-dependent Rabi frequency
\begin{equation}
    H \approx \frac{\hbar}{2} \left( \Omega(x^\textrm{eq}) + \Omega^\prime(x^\textrm{eq}) \, \delta x \right) e^{-i \Delta t}\sigma_+  +  \mathrm{H.c.},
\end{equation}
where the desired SDF coupling between motional and qubit degrees of freedom is generated by the derivative $\Omega^\prime(x)$ around the classical equilibrium positions. 

Let us now consider a chain of $N$ trapped ions in a linear Paul trap~\cite{PhysRevA.45.6493}. As we mentioned above, the transversely-addressed gradient interaction produces a position-dependent Rabi frequency under a combination of transversal Hermite-Gauss modes. Using different setups, this can be achieved by the generation of either an intensity or a polarization gradient for each ion along the trap-axis direction~\cite{faorlin2026entanglingionsengineeredlight, Cui2025}. In both cases, the mechanism enabling such gates is the same as the one responsible for the standard implementation of the LS gate, i.e. the differential AC Stark shift of energy levels experienced by the ions~\cite{PhysRev.100.703}. The AC Stark shift accounts for the response of an atom interacting with an oscillating electric field. Assuming a combination of laser fields $\boldsymbol{E}(t)=\frac{1}{2}\sum_j\mathcal{E}_je^{-i\omega_{L_j}t}\boldsymbol{e}_j+\mathrm{H.c.}$ and within the dipole approximation, the energy of a two-level system with frequency $\omega_0$ experiences a shift described by~\cite{PhysRevA.109.052417,PhysRevA.93.063404, Le_Kien_2013}
\begin{equation}
\label{eq: ac-stark_shift}
\delta E = -\frac{1}{4}\sum_{j}\Re{\frac{\Omega_j^2}{\delta_j}+\sum_{i>j}\Omega_i\Omega_j^*\left(\frac{1}{\delta_i}+\frac{1}{\delta_j}\right)\ee^{-\ii\Delta\omega_{ji} t}},
\end{equation}
where $\delta_j=\omega_0-\omega_{L_j}$ and $\Delta\omega_{ji}=\omega_{L_j}-\omega_{L_i}$. Note that we have explicitly incorporated the Rabi frequencies $\Omega_j=\mathcal{E}_j|\bra{0}\boldsymbol{e_j}.\boldsymbol{d}\ket{1}|$~\cite{Scully_Zubairy_1997}, with $\boldsymbol{d}$ the dipole operator of the atom~\cite{Scully_Zubairy_1997,  Le_Kien_2013}. We have assumed no spontaneous decay and, by considering $\Omega_j^2\ll\omega_0\approx\omega_{L_j}$, we neglected the off-resonant contributions proportional to $1/(\omega_0+\omega_{L_j})$. Equation~\eqref{eq: ac-stark_shift} accounts for constant single-beam shifts $\delta E_C=-\frac{1}{4}\sum_{j}\Re{\Omega_j^2/\delta_j}$, as well as dynamic two-beam ones, $\delta E_D(t)=-\frac{1}{4}\sum_{j}\Re{\sum_{i>j}\Omega_i\Omega_j^*\left(\frac{1}{\delta_i}+\frac{1}{\delta_j}\right)\ee^{-\ii\Delta\omega_{ji} t}}$. These shifts arise as a result of virtual transitions of the qubits' state, where a photon is either absorbed and emitted from the same laser mode, or absorbed and emitted from different laser modes, respectively~\cite{PhysRevLett.97.050505, PhysRevA.109.052417}. The crossed beam AC Stark shift is responsible for the state-dependent dipole forces, and for a combination of two laser beams, reads as
\begin{equation}
\label{eq: crossed_beam_shift}
\delta E_D(t)=-\frac{1}{8}\Omega_1\Omega_2^*\left(\frac{1}{\delta_1}+\frac{1}{\delta_2}\right)\ee^{-\ii\Delta\omega_L t}+\mathrm{H.c.},
\end{equation}
with $\Delta\omega_L=\omega_{L_1}-\omega_{L_2}$ being the laser beat-note. Let us assume that the coupling of spin and motional degrees of freedom is obtained via the spatial shaping of the Rabi frequencies, $\Omega_j\coloneq \Omega_j(\boldsymbol{r}_j)$. 
One way to produce the desired approximate $\mathrm{TEM}_{01}$ mode is through the use of a phase plate. In the following, instead of assuming the ideal $\mathrm{TEM}_{01}$, we make use of the electric field amplitude produced by a phase-plate. 
We first consider a Gaussian beam $\boldsymbol{E}_1(\boldsymbol{r}, t)$ with beam waist $W_0$ and Rayleigh range $z_0$~\cite{doi:https://doi.org/10.1002/0471213748.ch3}, which travels in the positive $z$ direction. This beam is then approximately focused at the focal plane of a lens, $z=f$, where the ions are located. Here we have assumed that the beam rays reaching the lens are paraxial~\cite{doi:https://doi.org/10.1002/0471213748.ch4}. The Fresnel approximation establishes a criteria under which, for free space, the phase accumulated by a traveling beam is approximately quadratic in the spatial frequencies $\nu_{\alpha}=k_{\alpha}/2\pi$, where, assuming that the beam travels in the $z$ direction, $\alpha=x,y$~\cite{doi:https://doi.org/10.1002/0471213748.ch4}. Under such approximation, the Fourier transform property of a lens~\cite{doi:https://doi.org/10.1002/0471213748.ch4, Goodman1969IntroductionTF} predicts the complex amplitude
\begin{equation}
U_{1}=\ii\mathcal{U}\frac{z_0}{f}\ee^{-\ii k_1(d+f)}\ee^{\ii\pi\frac{d-f}{\lambda_1 f^2}\rho^2}\ee^{-\frac{1}{{W'_0}^2}\rho^2},
\end{equation}
where $k_1=2\pi/\lambda_{L_1}$ is the wave-vector relation to the light wavelength, and $\rho^2=x^2+y^2$ is the radial distance to the beam axis. Here, $W_0'=W_0\frac{f}{z_0}$ denotes the output beam waist at the focal plane, and $d$ the distance traveled by the beam before reaching the lens. A schematic description of this process can be found in Fig.~\ref{fig: optical_scheme}\textbf{a}. 

In addition, we consider that a beam with an odd spatial profile is produced by passing the Gaussian beam through a phase plate. The transformed beam is then also focused by the lens, giving the complex amplitude
\begin{equation}
U_{2}=\frac{\mathcal{U}}{\sqrt{\pi}}\frac{z_0}{f}\ee^{-\ii k_2(d+f)}\ee^{\ii\pi\frac{d-f}{\lambda_2 f^2}\rho^2}\ee^{-\frac{1}{{W'_0}^2}y^2}D_+\left(\frac{x}{W'_0}\right),
\end{equation}
where we introduced the Dawson function $D_+(x)$~\cite{10.1119/1.15378}
\begin{equation}
\label{eq: dawson}
D_+(x)=\frac{1}{2}\int_0^\infty \mathrm{d}t \ee^{-\frac{t^2}{4}}\sin(xt).
\end{equation}

We note that this complex amplitude approximately resembles that of a $\textrm{TEM}_{10}$ mode. A comparison between complex amplitudes of the different TEM modes is depicted in Fig.~\ref{fig: optical_scheme}\textbf{b}.

Combining these amplitudes, we find the desired product in equation~(\ref{eq: crossed_beam_shift}), $\Omega\coloneq\Omega_1\Omega_2^{*}$, which reads
\begin{equation}\label{eq:rabi_freq_exp}
\Omega=\ii\Omega_L\ee^{-2\frac{1}{{W'_0}^2}y^2}\ee^{-\frac{1}{{W'_0}^2}x^2}D_+\left(\frac{x}{W'_0}\right),
\end{equation}
and where
\begin{equation}
\Omega_L=\frac{z_0^2|\mathcal{U}|^2}{f^2\sqrt{\pi}}\bra{0}\boldsymbol{e_1}.\boldsymbol{d}\ket{1}\bra{1}(\boldsymbol{e_2}.\boldsymbol{d})^*\ket{0}e^{i\psi^{\pm}(\rho)},
\end{equation}
which depends on a position-dependent phase $\psi^{\pm}(\rho)$ that reads 
\begin{equation}
\label{eq: pos_dependent_phase}
\psi^{\pm}(\rho)=\pi\left(\frac{1}{\lambda_{L_1}}\pm\frac{1}{\lambda_{L_2}}\right)\left(-2(d+f) + \frac{d-f}{f^2}\rho^2\right).
\end{equation}

For the laser wavelengths considered here, which are on the order of hundreds of nanometers, this phase can be safely neglected. A quantitative analysis justifying this approximation will be given in the next section. The $\pm$ sign depends on the chosen beam geometry, also discussed in the next section. 

\subsection{Ideal transversely-addressing light gradient gate}

The vibrational modes describing the collective motion of the ion chain about its classical equilibrium positions are described through the quantum mechanical operators $\delta \hat{x}_j= \sum_n \frac{M_{jn}^x}{\sqrt{2m\nu_n^x}}(a_n^x+{a_n^x}^{\dagger})$~\cite{_A_URA_2002}, where $M_{jn}^\alpha$ denotes the $j$-th coordinate of the $n$-th vibrational mode along the $\alpha$-axis~\cite{_A_URA_2002, Lee_2005}. As in our description of geometric phase gates, let us assume for the ideal gate description that there is only one relevant vibrational mode. We also assume that it belongs to the axial direction. For notational simplicity, we omit again the subscripts and superscripts for the selected mode, such that $\nu_{0}^x\coloneq\nu$, $M_{j0}^x\coloneq M_j$ and $a_0^x\coloneq a$.
A Taylor expansion of \eqref{eq:rabi_freq_exp} yields
\begin{equation}
\label{eq: rabi_expansion}
\hat{\Omega}(a^{\dagger},a)=\ii\Omega_L \eta M_{0}(a+a^{\dagger}) + \mathcal{O}(\eta^3),
\end{equation}
where 
\begin{equation}\label{eq:ld_parameter}
    \eta=\frac{1}{W_0'}\frac{1}{\sqrt{2m\nu}}
\end{equation} 
is an effective Lamb-Dicke (LD) parameter already introduced in Ref.~\citenum{Stopp2022}. The effective LD parameter plays the role of the usual LD parameter~\cite{RevModPhys.75.281}, but compares the spatial extent of the ground-state (GS) wavefunction with the beam waist $W_0^\prime$ at the focal plane, rather than with the laser wavelength $\lambda$.

We consider the regime in which the oscillator's GS width
around its equilibrium position is small compared to the output
beam waist. In this limit, the rotated-frame Hamiltonian
describing the transversely addressed gradient interaction
is well approximated by
\begin{equation}
\label{eq: eff_hamiltonian}
\begin{split}
H_{TG}=i\frac{1}{2}&\mathcal{F}_{\mathrm{TG}}(\tilde{a}+\tilde{a}^{\dagger})\ee^{\ii\Delta\omega_Lt}(Z_{j_1}+Z_{j_2})+\mathrm{H.c.},\\
\mathcal{F}_{\mathrm{TG}}&=-\frac{1}{8}\eta\Omega_L\left(\frac{1}{\delta_1}+\frac{1}{\delta_2}\right),
\end{split}
\end{equation}
where $\tilde{a}=ae^{-i\nu t}$ and
$\tilde{a}^{\dagger}=a^{\dagger}e^{i\nu t}$ are
rotated operators, and $(j_1,j_2)$ denotes the ion
pair involved in the gate. When the detuning of the laser beat-note frequency with respect to the motional mode frequency fulfills $\Omega_L\ll\Delta\omega_L\approx\nu$, in view of the rotating wave approximation~\cite{Scully_Zubairy_1997}, the off-resonant terms oscillating with the sum of frequencies can be ignored. Hence, the resulting Hamiltonian (\ref{eq: eff_hamiltonian}) may indeed be employed to approximate equation (\ref{eq: sd_force}) in the Pauli-$z$ basis.

\section{Motional error budget}
\label{sec. error_analysis}
In this section, we quantify the main errors arising from the imperfect evolution generated by the expansion terms and motional modes neglected in~\eqref{eq: eff_hamiltonian}. We first analyze the infidelity contribution from the disregarded terms in the dynamics of \textit{active} modes, i.e. those purposefully addressed to drive the gate. Then, we extend our study to account for the contribution of \textit{spectator} modes which, despite being spectrally well separated from the active modes, can generate residual entanglement between the spin and phonon subspaces, constituting a fundamental source of gate errors that must be carefully accounted for.

\subsection{Infidelity contribution from the active mode}
\label{subsec. fidelity1}
Let us first start by considering only one vibrational mode acting as a data bus~\cite{_A_URA_2002}. For a quantitative estimate of this vibrational mode's contribution to gate infidelity, we consider the higher-order terms in the Taylor expansion reported in~\eqref{eq: rabi_expansion} \cite{Kirchhoff2025}. Let us again assume the minimal gate description in which the spins are coupled to an axial mode in the trap axis direction $x$, acting as the mentioned active mode.
Additionally, we consider a position-dependent Rabi frequency $\Omega(x)$, which is an odd function with respect to the spin positions. In the rotated frame with respect to $H_0$, a Taylor expansion of such frequency yields
\begin{equation}
\label{eq: expanded_hamiltonian}
H=\ii\frac{\mathcal{F}}{2}(Z_{j_1}+Z_{j_2})\sum_{\substack{n\\  \mathrm{odd}}}c_n\eta^{n-1}\sum_{n'\leq n}\tau(n,n')(\ee^{\ii\varphi_{n,n'}^-t}-\ee^{\ii\varphi_{n,n'}^+t}),
\end{equation}
where $c_n$ are coefficients that depend on the specific spatial profile of $\Omega(x)$. One power of $\eta$ was absorbed into $\mathcal{F}$, which was redefined for convenience. We denote by $\tau(n,n')$ the operator accounting for all possible combinations of boson operator strings of length $n$, containing $n^\prime$ creation operators $a^{\dagger}$,
\begin{equation}
\tau(n,n')=\sum_{\mathrm{sym}}(a^{\dagger})^{n'}a^{n-n'}.
\end{equation}
We also introduced
\begin{equation}
\label{eq: rotating_frecs}
\varphi_{n,n'}^{+}=\delta l+\Sigma(l+1);\hspace{3ex}\varphi_{n,n'}^-=\Sigma l+\delta(l+1),
\end{equation}
with $l=n'-\frac{n+1}{2}$. We implicitly introduced the detuning $\delta=\nu-\nu_d$ and the sum of frequencies $\Sigma=\nu+\nu_d$, where $\nu_d$ is a driving frequency which, for our chosen realization in~\eqref{eq: eff_hamiltonian}, coincides with the laser beat-note $\nu_d=\Delta\omega_L$. The functions~(\ref{eq: rotating_frecs}) quantify how `resonant' each summand of expression~(\ref{eq: expanded_hamiltonian}) is, understanding by resonant those contributions that rotate with the detuning. Thus, only $l=-1,0$ provide resonant contributions, whereas any other value yields off-resonant ones. 

The ideal $ZZ$ gate induced by~\eqref{eq: sd_force} arises from the leading-order term in $\eta$ and the resonant contribution to~\eqref{eq: expanded_hamiltonian}. All other terms, either off-resonant or higher order in $\eta$, will contribute to the infidelity of the actual gate. 
Having identified these higher order in $\eta$ contributions, we now consider the Magnus expansion~\cite{https://doi.org/10.1002/cpa.3160070404} of the time evolution operator induced by~\eqref{eq: expanded_hamiltonian}.
For the evaluation of leading-order corrections to the ideal gate, we keep track of all resonant terms in~\eqref{eq: expanded_hamiltonian} up to quadratic order, whereas we only keep the constant terms for the less relevant off-resonant contributions. We note that, by retaining these higher-order terms in $\eta$ of the interaction picture Hamiltonian, the Magnus expansion now terminates at the 4th-order term, leading to

\begin{equation}
\label{eq: gate}
U(t)=\ee^{\alpha(t)(Z_{j_1}+Z_{j_2})+\mathcal{J}(t)(\mathbb{1}+Z_{j_1}Z_{j_2})}=\sum_{\boldsymbol{\sigma}}\mathcal{U}_{\boldsymbol{\sigma}}(t)\ket{\boldsymbol{\sigma}}\bra{\boldsymbol{\sigma}},
\end{equation}
where $\boldsymbol{\sigma}=(\sigma_1,\sigma_2)$ denotes the two-qubit computational basis and 
\begin{equation}
\begin{split}
\mathcal{U}_{00}(t)&=\ee^{2(\mathcal{J}(t)+\alpha(t))},\hspace{1ex}\mathcal{U}_{11}(t)=\ee^{2(\mathcal{J}(t)-\alpha(t))}, \\&\hspace{4ex}\mathcal{U}_{01}(t)=\mathcal{U}_{10}(t)=1.
\end{split}
\end{equation}

The functions $\mathcal{J}(t)\coloneq\mathcal{J}(t;\hspace{1ex}a^{\dagger},a)$ and $\alpha(t)\coloneq\alpha(t;\hspace{1ex}a^{\dagger},a)$ contain the contributions from the even and odd orders of the Magnus expansion, respectively:
\begin{equation}
\label{eq: magnus_functions}
\begin{split}
\mathcal{J}(t)&=\mathcal{J}_0(t)+\mathcal{J}_1(t)a^{\dagger}a,\\
\alpha(t)=a^{\dagger}&\big(\alpha_0(t)+\alpha_1(t) a^{\dagger}a\big) - \mathrm{H.c.}.
\end{split}
\end{equation}
The explicit expressions for the functions $\mathcal{J}_0(t)$, $\mathcal{J}_1(t)$, $\alpha_0(t)$ and $\alpha_1(t)$ are given in Appendix~\ref{appendix 1}.

We estimate the gate error by computing the average gate fidelity $\bar{F}(U,\mathcal{E})$~\cite{NIELSEN2002249} of the gate described by ~\eqref{eq: gate}, with respect to the ideal $ZZ$ gate. Given a noisy implementation of a unitary gate $U$ denoted by the quantum channel $\mathcal{E}$, the average gate fidelity is defined by~\cite{NIELSEN2002249}
\begin{equation}
\bar{F}(U,\mathcal{E})=\int\mathrm{d}\psi\bra{\psi}U^{\dagger}\mathcal{E}(\psi)U\ket{\psi},
\end{equation}
where the integral is over the Haar measure and $\psi=\ketbra{\psi}{\psi}$. It is convenient to express this quantity in terms of the entanglement fidelity as~\cite{Horodecki1999}
\begin{equation}
\bar{F}(U,\mathcal{E})=\frac{dF_{\mathrm{e}}(U^{-1}\circ\mathcal{E})+1}{d+1},
\end{equation}
with $d=\mathrm{dim}\{\mathcal{H}\}$ and where the entanglement fidelity is defined as \cite{Schumacher1996}
\begin{equation}
F_{\mathrm{e}}(\mathcal{E})=\bra{\Phi}(\mathbb{1}\otimes\mathcal{E})(\Phi)\ket{\Phi}.
\end{equation}

Here $\ket{\Phi}=\sum_{\boldsymbol{\sigma}}\ket{\boldsymbol{\sigma}}\otimes\ket{\boldsymbol{\sigma}}/\sqrt{d}$ refers to the maximally entangled state on a doubled Hilbert  space $\mathcal{H}\otimes \mathcal{H}$~\cite{Nielsen_Chuang_2010}. Assuming a separable initial state $\rho=\rho_{s} \otimes \rho_\mathrm{ph}$ in which $\rho_{\mathrm{ph}}$ denotes the phononic part, an ideal unitary channel $U_{\mathrm{id}}=\exp(-\ii \frac{\pi}{4} Z_1 Z_2)$ and the gate channel~\eqref{eq: gate} $\mathcal{E}=U(t)$, we find
\begin{equation}
\label{eq: fidelity_e}
\begin{split}
&F_{\mathrm{e}}(U_{\mathrm{id}}^{-1}\circ U(t))=\frac{1}{16}\sum_{\boldsymbol{\sigma,\sigma'}}\Lambda_{\boldsymbol{\sigma,\sigma'}}\mathrm{tr}\{\mathcal{U}_{\boldsymbol{\sigma'}}^{\dagger}(t)\mathcal{U}_{\boldsymbol{\sigma}}(t)\rho_{\mathrm{ph}}\},\\
&\Lambda_{\boldsymbol{\sigma,\sigma'}}=\frac{1}{2}(1+\varepsilon_{\boldsymbol{\sigma}}\varepsilon_{\boldsymbol{\sigma'}}+\ii(\varepsilon_{\boldsymbol{\sigma}}-\varepsilon_{\boldsymbol{\sigma'}})), \hspace{2ex}\varepsilon_{\boldsymbol{\sigma}}=(-1)^{\delta_{\sigma_1,\sigma_2}}.
\end{split}
\end{equation}

Here, $\mathcal U_{\boldsymbol{\sigma}}$ are the projections of channel $\mathcal E$ onto the computational basis states. Alternatively, for the purpose of benchmarking our results against experimental data, we may employ the state fidelity $\mathrm{F}=\braket{\psi_0\lvert U^{\dagger}\mathcal{E}(\rho_0)U\lvert\psi_0}$~\cite{Nielsen_Chuang_2010}, given a target state $U\ket{\psi_0}$ and the initial one $\rho_0$. 
Since the representation of $U(t)$ in \eqref{eq: gate} is diagonal in the computational basis, and considering the particular case of $\ket{\psi_0}=\ket{++}$ and an initial state of the form $\rho_0=|\psi_0\rangle  \langle\psi_0| \otimes \rho_\mathrm{ph}$, the state fidelity and the entanglement fidelity in \eqref{eq: fidelity_e} coincide, $F=F_\mathrm{e}$. More generally, we prove in Appendix~\ref{appendix 2} that the sufficient conditions for this to happen are that the initial state is chosen to be the homogeneous superposition $|\psi_0\rangle=\frac{1}{\sqrt{d}}\sum_{\sigma}\ket{\sigma}$, and that the noise channel $U^{-1}\circ\mathcal{E}$ is a generalized dephasing channel~\cite{Wilde_2016}.

We then proceed to obtain values for the average gate infidelity $1-\bar{F}$ in terms of the physical parameters describing the interaction: $\eta, \delta, t$ and $\mathcal{F}$. While the ideal gate motivates the choice $t_g=\frac{2\pi}{\delta}$ and $\mathcal{F}=\frac{\delta}{2}$, selecting the force such that $\mathcal{J}_0(t_g)=\frac{\pi}{4}$ returns better fidelity values by incorporating off-resonant and second-order contributions to the spin-spin coupling. We note that there exist other $t$ and $\mathcal{F}$ combinations which lead to an even greater fidelity. Nevertheless, this condition is difficult to predict in advance, as second-order contributions to $\mathcal{F}$ and $\alpha$ introduce a dependence of fidelity on the motional state of the ions.

Let us finally note that geometric-phase gates are sometimes implemented in conjunction with a Hahn-echo \cite{Hahn1950} pulse, which cancels out errors stemming from unequal illumination of the ions and unclosed phase-space trajectories~\cite{Kim2008, PhysRevLett.109.020503}. This corresponds to an evolution described by the sequence
\begin{equation}
\label{eq: spin_echo}
U_{\mathrm{echo}}(t)=U^{(2)}(X_{j_1}X_{j_2}\otimes\mathbb{1}_{\mathrm{ph}})U^{(1)}=\sum_{\boldsymbol{\sigma}}\mathcal{U}_{\bar{\boldsymbol{\sigma}}}^{(2)}\mathcal{U}_{\boldsymbol{\sigma}}^{(1)}\ket{\bar{\boldsymbol{\sigma}}}\bra{\boldsymbol{\sigma}},
\end{equation}
where $\bar{\boldsymbol{\sigma}}$ denotes the complementary of the bit-string $\boldsymbol{\sigma}$, and now $t_{g,\mathrm{echo}}=2t_g$, $\mathcal{F}_{\mathrm{echo}}=\mathcal{F}/\sqrt{2}$. By $U^{(1)}$, $U^{(2)}$ we refer to the action of the unitary in~\eqref{eq: gate} during the time intervals $(0,t/2)$ and $(t/2,t)$, respectively. The products in~\eqref{eq: spin_echo} are
\begin{equation}
\begin{split}
\label{eq: projections_products}
&\mathcal{U_{00}}^{(2)}\mathcal{U_{11}}^{(1)}\approx\ee^{2\tilde{\mathcal{J}}+2\Delta\alpha_0}\ee^{-\ii2\varphi},\\
&\mathcal{U_{11}}^{(2)}\mathcal{U_{00}}^{(1)}\approx\ee^{2\tilde{\mathcal{J}}-2\Delta\alpha_0}\ee^{-\ii2\varphi},\\
&\mathcal{U_{01}}^{(2)}\mathcal{U_{10}}^{(1)}=\mathcal{U_{10}}^{(2)}\mathcal{U_{01}}^{(1)}=\mathbb{1},
\end{split}
\end{equation}
where $\tilde{\mathcal{J}}=\mathcal{J}^{(1)}+\mathcal{J}^{(2)}$ and $\Delta\alpha_0=\alpha_0^{(2)}-{\alpha_0^{(1)}}$. For the derivation of the approximate exponents, we invoked the Baker-Campbell-Hausdorff (BCH) formula~\cite{Hall2015} and found no second-order resonant contributions arising from the commutators, while the only first-order off-resonant contribution is the complex number $-\ii2\varphi$, with $\varphi = 2\Im{\alpha_0^{(2)}({\alpha_0^{(1)}})^*}$. This approximate quantum channel allows us to analytically obtain the expectation values appearing in~\eqref{eq: fidelity_e} when the initial phonon states are assumed to follow a thermal distribution $\rho_{\mathrm{ph}}=\sum_{n}p_{n}\ket{n}\bra{n}$, where $p_{n}=\ee^{-\beta\nu n}(1-\ee^{-\beta\nu})$ \cite{gardiner2004quantum}. This is possible because the off-diagonal structure in~\eqref{eq: spin_echo} does not change the form of the entanglement fidelity~(\ref{eq: fidelity_e}), provided the identification $\mathcal{U}_{\boldsymbol{\sigma}}=\mathcal{U}_{\boldsymbol{\bar{\sigma}}}^{(2)}\mathcal{U}_{\boldsymbol{\sigma}}^{(1)}$.

Also, one needs to fix the Taylor expansion coefficients in expression~(\ref{eq: expanded_hamiltonian}). Motivated again by the experimental implementation described in Subsec.~\ref{subsec. implementation}, we assumed a Rabi frequency with the form 
\begin{equation}
\label{physical_rabi}
\hat{\Omega}(a^{\dagger},a)=\mathcal{F}\ee^{-\left(\eta(a+a^{\dagger})\right)^2}D_+(\eta(a+a^{\dagger})).
\end{equation}

We note that the previous formalism is general and can be used to analyze other experimental realizations. It follows that the average gate fidelity reads
\begin{equation}
\label{eq: fidelity_analytic}
\bar{F}=\frac{1}{10}\left(5+ \lim_{\beta\to0} E_{\bar{n}}(0,\beta,\Delta\alpha_0)+4\Im{\ee^{\ii2\varphi}E_{\bar{n}}(\tilde{\mathcal{J}}_0,\tilde{\mathcal{J}}_1,\Delta\alpha_0)}\right),
\end{equation}
where we find
\begin{widetext}
\begin{equation}
\label{eq: expectations}
E_{\bar{n}}(\alpha,\beta,\gamma)=\frac{\ee^{\alpha}}{1+\bar{n}(1-\ee^{\beta})}\exp{-\frac{|\gamma|^2}{\beta^2}\left(\ee^{\beta}-\beta-1+\frac{\bar{n}}{1+\bar{n}(1-\ee^{\beta})}\left(\ee^{\beta}-1\right)^2\right)}.
\end{equation}
\end{widetext}

\begin{figure}
  \centering \includegraphics[width=1\columnwidth]{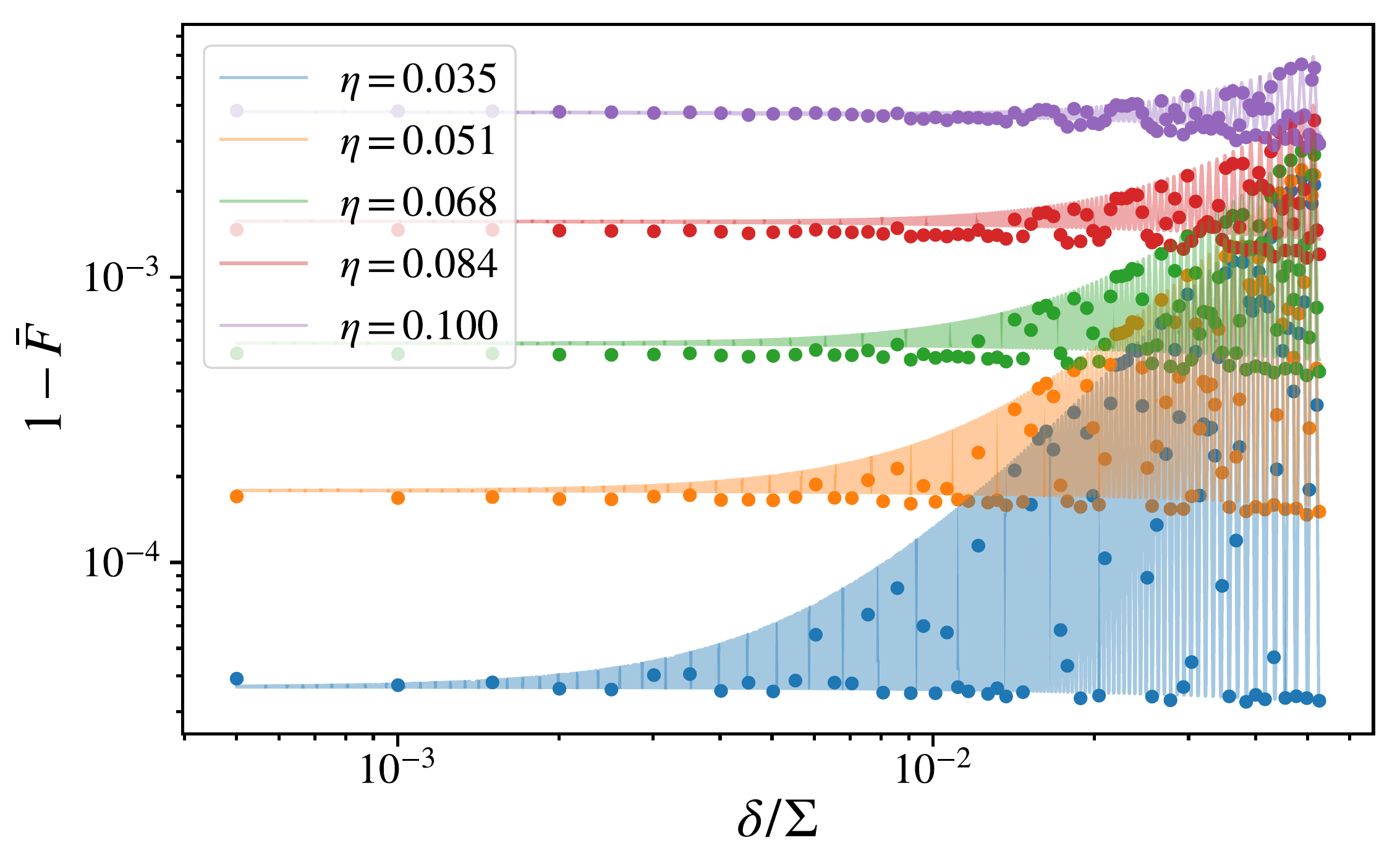}
  \caption{\textbf{Average gate infidelity due to motional errors:} 
  We consider here only the active mode, but include  higher-order terms of the LD expansion for different values of the LD parameter, as a function of $\delta/\Sigma$.
  We benchmark the analytical results given by Eq.~(\ref{eq: fidelity_analytic}) (solid lines) against those obtained by the computation of the approximate propagator produced by the non-expanded version of the Hamiltonian~\eqref{eq: expanded_hamiltonian} (data points). For the numerical estimation, we set a cutoff for the dimension of the phonon subspace, $\textrm{dim}(\mathcal{H}_{p})=15$. We set $\bar{n}=0.1$. We sweep values in the $\delta/\Sigma$ axis by fixing the axial trap frequency $\nu$ and changing the laser beat-note $\Delta\omega_\mathrm{L}$, such that the LD parameter remains constant with the change of $\delta/\Sigma$. For improved visualization in this figure, we do not employ physical parameter values this time.}
  \label{fig: f_vs_delta}
\end{figure}

By evaluating expression~(\ref{eq: fidelity_analytic}) for different $\eta$ and $\delta$ values, we obtain Fig.~\ref{fig: f_vs_delta}. The ratio $\delta/\Sigma$ accounts for the relevance of the off-resonant contributions to the infidelity. We benchmark the analytical results obtained with Eq.~(\ref{eq: fidelity_analytic}) (solid lines) against those obtained by the computation of the approximate propagator produced by the non-expanded version of the Hamiltonian~\eqref{eq: expanded_hamiltonian} (data points). For the computation of such an operator, we set a cutoff dimension for the phonon subspace, for which we find it sufficient to set $\textrm{dim}(\mathcal{H}_{p})=15$. For this plot we use $\bar{n}=0.1$.  
Then, we use \textit{QuTiP}~\cite{qutip5} to numerically solve the Schrödinger equation $\ii\hbar,dU/dt=H(t)U$, subject to the initial condition $U(0)=\mathbb{1}$, where $\mathbb{1}$ denotes the identity operator. The resulting evolution operator is applied to an informationally complete set of initial states~\cite{Nielsen2021gatesettomography}.

Some minor discrepancies are observed, which we attribute to the terms in expansion~\eqref{eq: expanded_hamiltonian} not accounted for by the analytical calculation, as well as to the approximations made to obtain the products in~\eqref{eq: projections_products}. Given that the main purpose of this figure is to assess the validity of~\eqref{eq: fidelity_analytic} against the numerical fidelity estimations, we do not employ physical parameter values this time. We conduct physically-relevant fidelity estimations in the next subsection, in which we also account for the infidelity contributions stemming from the remaining motional modes of the crystal.

\subsection{Infidelity contribution from spectator modes}
\label{subsec. fidelity2}
 
For a chain with $N$ ions, the ideal time evolution operator mediates the interaction between qubit states through one -- or possibly several -- of the $3N$ vibrational modes, the active modes. The remaining phononic modes are regarded as spectator modes, and they become relevant when considering further deviations from the ideal gate evolution.
Aiming at the study of such deviations, we now assume $N$ motional modes in each direction. We will model gates acting on two out of the $N$ ions. 

Consider the dimensionless deviation $\delta \overline r$ of ion $j$ from its classical equilibrium position, with components $\delta \bar{x}_j$, $\delta \bar{y}_j$ and $\delta \bar{z}_j$, which we define by
\begin{equation}
\delta \bar{x}_j=\sum_{m_x}\eta_{m_x}^x M_{m_x,j}^x(a_{m_x}^x+{a_{m_x}^x}^{\dagger}),
\end{equation}
and equivalently in the other directions. The parameter $\eta_m^\alpha$
represents the counterpart of the LD parameter for direction $\alpha$ and mode $m$~\cite{Zhu2006}.

Logical operations implemented on ion traps are susceptible to crosstalk errors, originating from residual illumination of neighboring ions not involved in the gate \cite{ParradoRodriguez2021}. 
Particularly, when the approximated $\mathrm{TEM}_{01}$ mode is generated through the phase-plate, it may show large deviations from the target Laguerre-Gauss amplitude, exhibiting longer range tails which can induce crosstalk with neighboring ions \cite{Mai2026}.
In general, the exact alignment of the $\textrm{TEM}_{01}$ gradient will be tilted by an angle $\phi$ with respect to the ion chain axis. This contributes to reduce the effect of the aforementioned crosstalk.
The coupling Rabi frequency is then written in terms of new coordinates $\xi_j^\alpha$ 
\begin{equation}
\begin{split}
&\delta \bar{x}_j\to\xi^x_j=\delta \bar{x}_j\cos\phi +\delta \bar{y}_j\sin\phi \\
&\delta \bar{y}_j\to\xi^y_j=-\delta \bar{x}_j\sin\phi +\delta \bar{y}_j\cos\phi.
\end{split}
\end{equation}

The equivalent of expression~(\ref{eq: rabi_expansion}) then reads $\hat{\Omega}_j=\ii\Omega_L\big( \xi^x_j+\epsilon_j(x,y,z)\big)$ where, up to third order in fluctuation operators, we have introduced
\begin{equation}
\label{eq: correction}
\epsilon_j(x,y,z)\simeq -i\delta \bar{z}_j\xi^x_j-\frac{5}{3}(\xi^x_j)^3-2\delta \bar{y}_j(\xi^x_j)^2-\xi^x_j(\delta \bar{z}_j)^2.
\end{equation}

We consider the effect that this rotation has on the thermal motional error, as it introduces a direct off-resonant coupling to other motional modes. We recall that the LD parameters in the directions transverse to the laser propagation are defined by 
\begin{equation}
\label{eq: ld_parameter_xy}
\eta_{n_{\alpha}}^{\alpha}=\frac{1}{W_0'}\frac{1}{\sqrt{2m\nu_{n_{\alpha}}^{\alpha}}}; \hspace{3ex} \alpha=x,y, 
\end{equation}
whereas for the $z$ direction, we find
\begin{equation}
\label{eq: ld_parameter_z}
{\eta_{n_z}^z}^{\pm}=2\pi\frac{1}{\sqrt{2m\nu_{n_z}^z}}\left(\frac{1}{\lambda_{L_1}}\pm\frac{1}{\lambda_{L_2}}\right).
\end{equation}

Here, ${\eta_{n_z}^z}^{+}$ and ${\eta_{n_z}^z}^{-}$ represent the recoil-induced LD parameters. They denote the parameters arising from a counter-propagating beam geometry and a co-propagating beam geometry, respectively. For the values of the laser wavelengths and the output beam waist in this work, both in the order of hundreds of nanometers, we note that ${\eta_{\mathrm{com}}^z}^{-}$ is orders of magnitude below $\eta^x_{\mathrm{com}_x}$ and $\eta^y_{\mathrm{com}_y}$. In contrast, ${\eta_{n_{\mathrm{com}_z}}^z}^{+}$ takes values similar or above those of the LD parameters transverse to propagation, becoming the main source of deviation from the ideal gate and making the counter-propagating geometry significantly more prone to couplings to the $z$-direction vibrational modes. 

Let us briefly revisit our analysis regarding the phase~\eqref{eq: pos_dependent_phase}, which we neglected previously without a quantitative calculation. Contrary to ${\eta_{{\mathrm{com}}_z}^z}^{+}$, the values of the position-dependent term of the phase $\psi_z^+\sim ({\eta_{{\mathrm{com}}_z}^z}^{+}/\sqrt{1/\lambda_{L_1}+1/\lambda_{L_2}})^2$ are not comparable to those of the LD parameters transverse to propagation, justifying our approximation.
Irrespective of the beam geometry, mirroring the analysis that we performed in Subsec.~\ref{subsec. fidelity1}, we incorporate the correction~(\ref{eq: correction}) to the active terms accounted for after Hamiltonian~(\ref{eq: expanded_hamiltonian}) and compute the resulting propagator. As previously stated, we keep resonant terms up to $(\eta^{\alpha}_{n_{\alpha}})^2$ order and off-resonant terms up to $(\eta^{\alpha}_{n_{\alpha}})$ order. Note that there are now linear terms which were not present in the single-mode analysis, participating in the evolution through the $z$-direction of motion. We obtain the operator
\begin{equation}
\label{eq: gate_complete}
U_{\mathrm{TG}}(t)=\ee^{\gamma(t)+\alpha_{j_1}(t)Z_{j_1}+\alpha_{j_2}(t)Z_{j_2}+\mathcal{J}(t)Z_{j_1}Z_{j_2}}=\sum_{\boldsymbol{\sigma}}\ket{\boldsymbol{\sigma}}\bra{\boldsymbol{\sigma}}\mathcal{U}_{\boldsymbol{\sigma}}(t),
\end{equation}
where $\mathcal{U}_{\boldsymbol{\sigma}}(t)=\ee^{\Theta_{\boldsymbol{\sigma}}(t)}$ and
\begin{equation}
\label{eq: projections_spectator}
\Theta_{\boldsymbol{\sigma}}(t)=\gamma(t)+(-1)^{\sigma_{j_1}}\alpha_{j_1}(t)+(-1)^{\sigma_{j_2}}\alpha_{j_2}(t)+(-1)^{\sigma_{j_1}+\sigma_{j_2}}\mathcal{J}(t).
\end{equation}

As reported above in~\eqref{eq: magnus_functions}, $\mathcal{J}(t)\coloneq\mathcal{J}(t;\hspace{1ex}\{{a^{\alpha}_{\alpha_n}}^{\dagger}\},\{a^{\alpha}_{\alpha_n}\})$ and $\gamma(t)\coloneq\gamma(t;\hspace{1ex}\{{a^{\alpha}_{\alpha_n}}^{\dagger}\},\{a^{\alpha}_{\alpha_n}\})$ are diagonal operators in the number basis, and stem from the even orders of the Magnus expansion. Contrary, the off-diagonal operators $\alpha_j(t)\coloneq\alpha_j(t;\hspace{1ex}\{{a^{\alpha}_{\alpha_n}}^{\dagger}\},\{a^{\alpha}_{\alpha_n}\})$ arise from odd orders. For consistency with our second-order-resonant criteria, we also compute the functions $\gamma(t)$ and $\alpha_j(t)$ up to fourth-order in the Magnus expansion. As we did for the single-mode version of the gate, we write these functions explicitly in Appendix~\ref{appendix 1}. 

\begin{figure*}[t!]
\centering
\includegraphics[width=\textwidth]{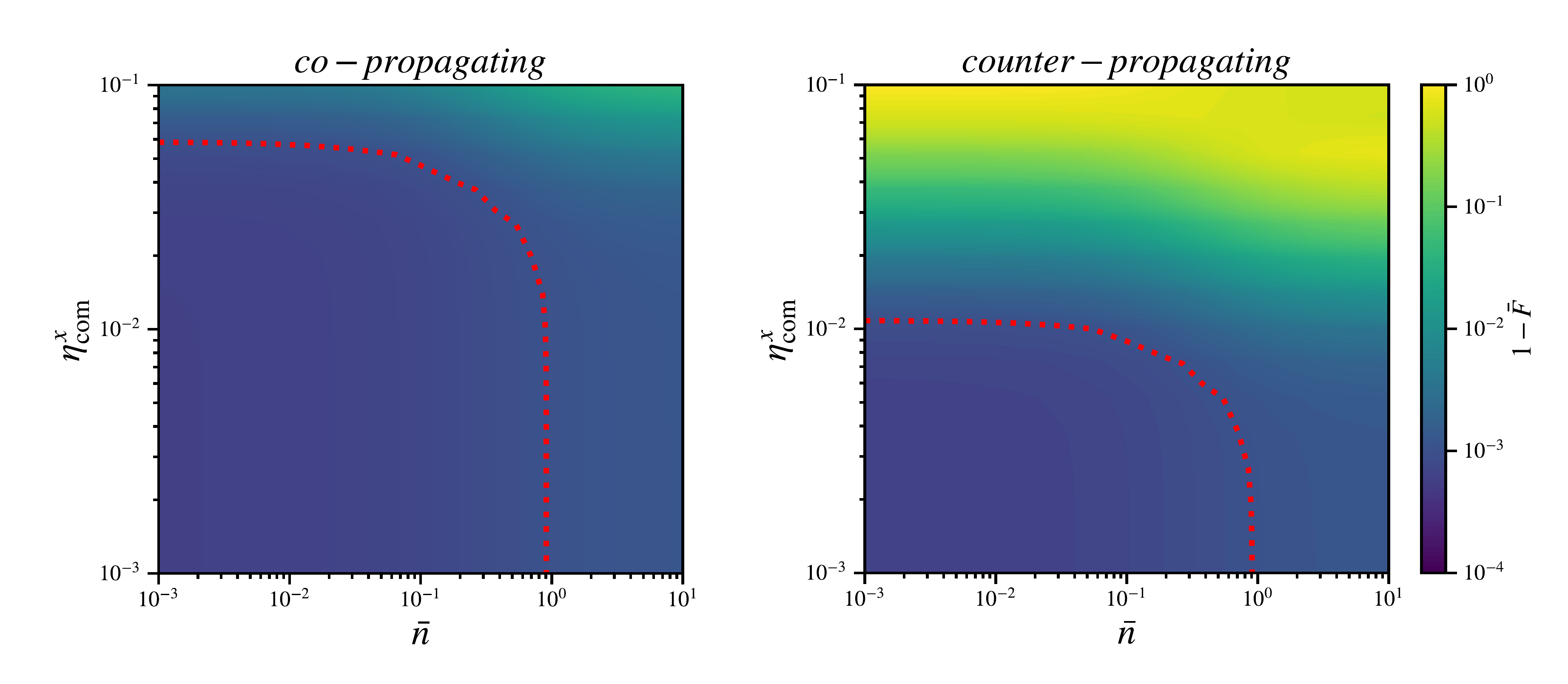}
\caption{\textbf{Motional-error contribution to the average gate infidelity for co-propagating and counter-propagating beam schemes:}.  We evaluate numerically expression~(\ref{eq: fidelity_e})  as a function of $\eta^x_{\mathrm{com}}$ and $\bar{n}$, making use of the evolution operator~(\ref{eq: projections_spectator}).  We obtain gate infidelity estimations with contributions both from active and spectator modes, assuming evolution under a Hahn-echo sequence. We assume a chain of two ions and that all motional modes are cooled up to the same average phonon number. Red dashed lines are used to separate the regions with infidelities under and above $10^{-3}$. Given a value for $\eta^x_{\mathrm{com}}$, we obtain the rest of the effective Lamb-Dicke parameters for the directions and modes according to Eqs.~(\ref{eq: ld_parameter_xy})~and~(\ref{eq: ld_parameter_z}). We use the values $(\omega^{(x)},\, \omega^{(y)},\, \omega^{(z)})=2\pi\times(0.3422,\, 1.344,\, 1.304)\,\mathrm{MHz}$ for the secular trap frequencies and a detuning $\delta = 2\pi \times 20\mathrm{kHz}$. We also set $N_{\mathrm{loops}}=4$. These results confirm that the co-propagating geometry strongly suppresses the infidelity contributions stemming from the occupation of collective modes in the $z$ direction.}
\label{fig: inf_countourplots}
\end{figure*}

We aim again at modeling a gate with a Hahn-echo pulse. In the more general scenario in which a gate is composed of an even number of loops in phase space $N_{\mathrm{loops}}$, and considering that the echo pulse is executed after the first half of them, we take the corresponding products of the projection operators in Eq.~(\ref{eq: projections_spectator}). Using the BCH formula, we find
\begin{equation}
\label{eq: spin_echo_general}
U_{\mathrm{echo}}^{N_{\mathrm{loops}}}(t)\approx\sum_{\boldsymbol{\sigma}}\mathcal{U}_{\boldsymbol{\sigma}}^{N_{\mathrm{loops}}}\ket{\bar{\boldsymbol{\sigma}}}\bra{\boldsymbol{\sigma}},
\end{equation}
with $\mathcal{U}_{\boldsymbol{\sigma}}^{N_{\mathrm{loops}}}=\exp\left(\sum_{l=0}^{N_{\mathrm{loops}}-1}{(\Theta_{\boldsymbol{\sigma}}^{(l)}+\sum_{l'<l}\theta_{\boldsymbol{\sigma}}^{(l,l')}})\right)$ and 
\begin{equation}
\begin{split}
\label{eq: phases_projector_general}
&\Theta_{\boldsymbol{\sigma}}^{(l)}=\gamma^{(l)}+s_l((-1)^{\sigma_{j_1}}\alpha_{j_1}^{(l)}+(-1)^{\sigma_{j_2}}\alpha_{j_2}^{(l)})+(-1)^{\sigma_{j_1}+\sigma_{j_2}}\mathcal{J}^{(l)},\\
&\theta_{\boldsymbol{\sigma}}^{(l,l')}=s_ls_{l'}\frac{1}{2}\comm{\alpha_{j_1}^{(l)}+(-1)^{\sigma_{j_1}\oplus\sigma_{j_2}}\alpha_{j_2}^{(l)}}{\alpha_{j_1}^{(l')}+(-1)^{\sigma_{j_1}\oplus\sigma_{j_2}}\alpha_{j_2}^{(l')}}.
\end{split}
\end{equation}
Here, we introduced $s_l=\mathrm{sign}\left({N_{\mathrm{loops}}/2-(l+1)}\right)$. Also, $\gamma^{(l)}=\gamma(t_l, t_{l+1})$ with $t_l=l(t/N_{\mathrm{loops}})$ --and equivalently for the other operators-- denote the intermediate evolutions for each of the $N_{\mathrm{loops}}$ loops comprising the gate. We substitute these expressions into Eq.~(\ref{eq: fidelity_e}) to obtain estimates for the average gate fidelity including contributions from spectator modes. We compute the expectation values in~\eqref{eq: fidelity_e} numerically, as we no longer find an analytical expression when the spectator modes are included. For this purpose, we again set a cutoff for the phonon-subspace. Fig.~\ref{fig: inf_countourplots} shows values for the average gate infidelity as a function of the mean thermal occupation number and the LD parameter of the COM axial mode. For these results, we assume that all motional modes are cooled down to the same average phonon number. We note, however, that cooling down spectator modes to a larger average phonon number does not lead to significant changes in fidelity. We compute infidelity values for two ion crystals both for the counter-propagating and co-propagating geometries, and for this case we set $N_{\mathrm{loops}}=4$.

\begin{figure}
\centering
\includegraphics[width=\columnwidth]{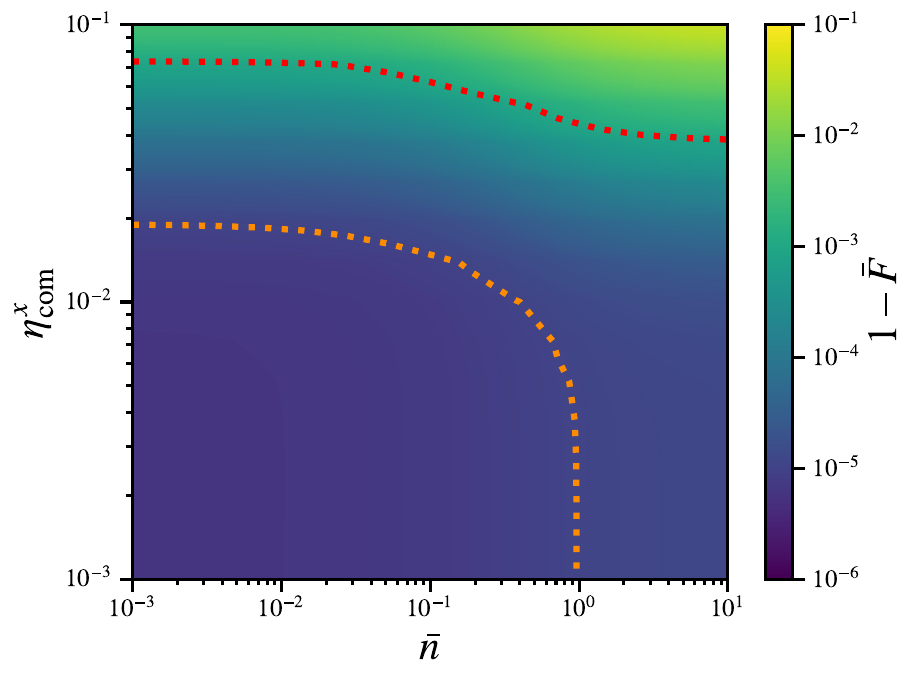}
\caption{\textbf{Spectator-mode motional error contribution to the average gate infidelity} We depict the average gate infidelity as a function of $\eta^x_{\mathrm{com}}$ and $\bar{n}$ for half the experimental detuning. For a chain of two ions and a co-propagating beam geometry, we obtain average gate infidelity estimations, evaluating expression~(\ref{eq: fidelity_e}) and making use of the evolution operator~(\ref{eq: projections_spectator}). All motional modes are assumed to be equally cooled. Red and orange dashed lines are used to separate the regions with infidelities under and above $10^{-3}$ and $10^{-5}$, respectively. Given a value for $\eta^x_{\mathrm{com}}$, we obtain the rest of the effective Lamb-Dicke parameters for the directions and modes according to Eqs.~(\ref{eq: ld_parameter_xy})~and~(\ref{eq: ld_parameter_z}). We use the values $(\omega^{(x)},\, \omega^{(y)},\, \omega^{(z)})=2\pi\times(0.3422,\, 1.344,\, 1.304)\,\mathrm{MHz}$ for the secular trap frequencies. We also set $N_{\mathrm{loops}}=4$. For this plot, we divide the experimental detuning employed in~\cite{faorlin2026entanglingionsengineeredlight} by a factor of two, $\delta = 2\pi \times 10\mathrm{kHz}$. In this way, we are able to predict a minimum infidelity estimate of, approximately, $5\times10^{-6}$.}
\label{fig: inf_countourplot_precise}
\end{figure}

The numerical results shown in Fig.~\ref{fig: inf_countourplots} confirm that the co-propagating geometry strongly suppresses the infidelity contributions stemming from the occupation of collective modes in the $z$ direction, as we anticipated in our discussion around ~\eqref{eq: ld_parameter_z}. This is the main motivation for the use of a co-propagating beam geometry, and explains why the counter-propagating geometry exhibits poor fidelity values. 
Consequently, we will assume a co-propagating beam geometry for the following calculations.
It is notable that the $z$ direction contributions to infidelity are smaller than the $x$ and $y$ second-order resonant contributions that typically characterize our expansion. As we will shortly show, neglecting this $z$ direction contributions does not lead to appreciable deviations from the numerical results, and thus we find it consistent to do so. The infidelity values in the co-propagating beam geometry plot shown in Fig.~\ref{fig: inf_countourplots} plateau at approximately $5\times10^{-4}$. This behavior arises because, for sufficiently small values of $\eta^x_{\mathrm{com}}$ and $\bar{n}$, leading-order off-resonant terms dominate the dynamics. Consequently, achieving higher gate fidelities requires reducing the detuning. For the investigation of this regime, we divide the experimental detuning used in Fig.~\ref{fig: inf_countourplots} by a factor of two, obtaining Fig.~\ref{fig: inf_countourplot_precise} following the same procedure described above. The resulting plot shows that infidelities as low as $5\times10^{-6}$ are within reach for the co-propagating beam geometry.

Finding an approximate description for the operator~\eqref{eq: spin_echo_general} with no $z$ mode dependence is advantageous, as it allows us to obtain a new analytical expression for infidelity in terms of expectation values of the form~(\ref{eq: expectations}) for each mode. Using the BCH formula to find the products in \eqref{eq: fidelity_e}, keeping corrections that satisfy the second-order resonant/first-order off-resonant criterion as we did in~\eqref{eq: projections_products}. As we can now analytically compute the infidelity values, we obtain estimations for longer ion chains. 
This becomes challenging to do numerically for more than a few ions: given a sufficiently accurate cutoff for the dimension of each phonon subspace, the dimension of the complete Hilbert space increases exponentially with the number of ions.

\begin{figure}
\centering
\includegraphics[width=\columnwidth]{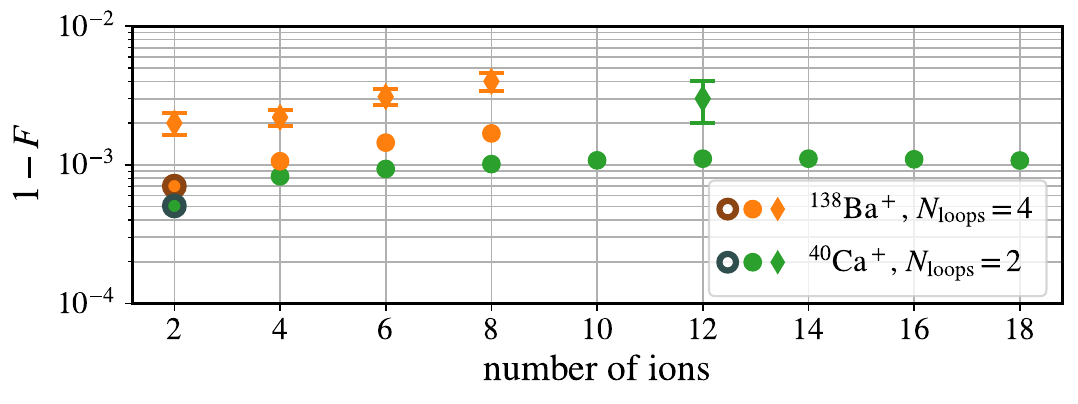}
\caption{\textbf{Comparison of the experimental state infidelity to the theoretical prediction with increasing ion number.} We evaluate expression~(\ref{eq: fidelity_e}) analytically (solid orange and green data-points) neglecting $z$ direction error contributions for a co-propagating beam geometry. In line with the experimental results shown in~\cite{faorlin2026entanglingionsengineeredlight}, we compute fidelity estimations for the $^{138}\mathrm{Ba}^+$ experiment up to 8 ions (orange data-points), with values $(\omega^{(x)},\, \omega^{(y)},\, \omega^{(z)})=2\pi\times(0.3422,\, 1.344,\, 1.304)\,\mathrm{MHz}$ for the secular trap frequencies, a detuning $\delta = 2\pi \times 20\mathrm{kHz}$ and $N_{\mathrm{loops}}=4$ loops in phase space. We do the same for the $^{40}\mathrm{Ca}^+$ experiment up to 18 ions (green data-point), with values $(\omega^{(x)},\, \omega^{(y)},\, \omega^{(z)})=2\pi\times(0.480,\, 4,\, 4)\,\mathrm{MHz}$ for the secular trap frequencies, a detuning $\delta = 2\pi \times 16.7\mathrm{kHz}$ and $N_{\mathrm{loops}}=2$ in phase space. For two ions, we also include the estimation via the numerical approximation of the expectation values in Eq.~(\ref{eq: fidelity_e}) (unfilled brown and dark green data-points for $^{138}\mathrm{Ba}^+$ and $^{40}\mathrm{Ca}^+$, respectively). We also plot the experimental data in Ref.~\citenum{faorlin2026entanglingionsengineeredlight} with diamond markers for comparison. In accordance with the experimental results, we predict high-fidelity ($>99.5\%$) entangling gates for medium-size ion crystals.}
\label{fig: inf_vs_ions}
\end{figure}

Figure~\ref{fig: inf_vs_ions} illustrates infidelity estimations as the ion-chain increases in size, given realistic values for the LD parameters and occupation numbers. This time, to reproduce the experimental setups in \cite{faorlin2026entanglingionsengineeredlight}, we set $N_{\mathrm{loops}}=4$ for the $^{138}\mathrm{Ba}^+$ experiment up to 8 ions (orange data-points), and $N_{\mathrm{loops}}=2$ for the $^{40}\mathrm{Ca}^+$ experiment up to 18 ions (green data-points). The maximum number of ions considered for each species is determined by the stability of the linear chain with the frequencies employed. We also plot the experimental data in Ref.~\citenum{faorlin2026entanglingionsengineeredlight} with diamond markers for comparison. For this plot, we make use of the state fidelity $F$ instead of the average gate fidelity $\bar{F}$. 

As can be appreciated from this figure, we predict high-fidelity entangling operations between pairs of ions for medium-size ion crystals (solid blue and green data-points). Each value is obtained by averaging the results for the innermost and outermost ion pairs. We also plot estimates of infidelity including $z-$deviations for two ions, computing the expectation values in Eq.~(\ref{eq: fidelity_e}) numerically (unfilled brown and dark green data-points for $^{138}\mathrm{Ba}^+$ and $^{40}\mathrm{Ca}^+$ respectively). Numerical estimates exhibit agreement with the analytical results, confirming that the $z$-direction contributions are not relevant for the co-propagating geometry. As evidenced by the simulation results for the $^{40}\mathrm{Ca}^+$ experiment, which is compatible with longer ion crystals, we appreciate a slight increase of infidelity with the length of the chain, in the transition from short to medium size crystals. This is because, with the increasing number of ions, more spectator modes appear and give rise to off-resonant contributions. Further from this initial region, the infidelity remains largely unaffected by increasing chain lengths, as the additional normal modes are very far from resonance.
We remark that in using \eqref{eq: spin_echo_general} for larger ion chains, we are considering that $\boldsymbol{\sigma}=(\sigma_1,\sigma_2)$ labels the projection onto the computational basis states corresponding only to the two ions $i,j$ involved in the gate, regardless of the number of ions in the chain. In particular, this means that within the results in Fig.~\ref{fig: inf_vs_ions} there is no crosstalk contribution to the infidelity.

The results in this figure are compatible with the experimental ones observed in Fig. 2c) of Ref.~\citenum{faorlin2026entanglingionsengineeredlight}. The difference observed with respect to such data is expected, and should be attributed to additional error sources affecting the experiment that are not considered in the analysis developed in this section. In fact, while we were also expecting to predict a slow degradation of fidelity with the increase of the chain size, the values obtained here are meant to be benchmarked against the spectator modes' contribution to the `error budget' shown in Tab. 1 of Ref.~\citenum{faorlin2026entanglingionsengineeredlight}, which the authors predict to be of $2.1\times10^{-4}$. We obtain a value of $\approx7\times10^{-4}$, and believe that the simulations performed in Ref.~\citenum{faorlin2026entanglingionsengineeredlight} slightly underestimate the error rate attributed to spectator modes.

In this section, we have assumed gates with fixed amplitudes. We note that, in practice, shaped pulses are commonly employed. As a consequence, the spectator mode's contribution to infidelity decreases with respect to the fixed amplitude case. We leave the modelling of shaped pulses as a line of research for future work. In the next section, we compute infidelity estimates which also consider qubit frequency fluctuations due to magnetic field noise.

\section{Dephasing error budget}
\label{sec: dephasing}
We finally extend our previous discussion to include dephasing in the fidelity calculations, accounting for stochastic fluctuations due to magnetic-field noise. We consider a stochastic process $\tilde{\delta}(t)$ that modifies the characteristic qubit frequency  $\omega_0\to\omega_0+\tilde{\delta}(t)$ \cite{van1992stochastic, gardiner1985handbook, gardiner2004quantum, PhysRevResearch.7.013008, Vinas2026ContextAwareNPJ}. In the rotated frame with respect to the unitary evolution~\eqref{eq: sd_evolution}, the stochastic Hamiltonian governing the action of such fluctuations reads
\begin{equation}
\tilde{H}(t)=\frac{1}{2}\tilde{\delta}(t)(Z_1+Z_2).
\end{equation}

Let us obtain an evolution under this Hamiltonian. We take the statistical average over stochastic trajectories of the density matrix, ${\rho}(t)=\mathbb{E}[\tilde{\rho}(t)]$. For an initial state $\rho_0$, the density matrix along a given trajectory is $\tilde{\rho}(t)=\ee^{-\ii\int_0^{t}\tilde{H}(t')\mathrm{d}t'}\rho_0\ee^{\ii\int_0^{t}\tilde{H}(t')\mathrm{d}t'}$. Such evolution results in the state 
\begin{equation}
\tilde{\rho}(t)=\sum_{\boldsymbol{\sigma},\boldsymbol{\sigma'}}\rho_{\boldsymbol{\sigma}, \boldsymbol{\sigma'}}(0)\ketbra{\boldsymbol{\sigma}}{\boldsymbol{\sigma'}}\ee^{-\ii\Upsilon_{\boldsymbol{\sigma},\boldsymbol{\sigma'}}\int_0^{t}\tilde{\delta}(t')\mathrm{d}t'},
\end{equation}
where 
\begin{equation}
\Upsilon_{\boldsymbol{\sigma},\boldsymbol{\sigma'}}=(-1)^{\sigma_{j_1}}+(-1)^{\sigma_{j_2}}-(-1)^{\sigma'_{j_1}}-(-1)^{\sigma'_{j_2}}.
\end{equation}

Noting that the commutation of the stochastic evolution operator by the refocusing pulse in the sequence~\eqref{eq: spin_echo} flips the sign of the Hamiltonian for the second half of the evolution, we find
\begin{equation}
\label{eq: dephasing_state}
\mathbb{E}[\tilde{\rho}(t_g)]=\sum_{\boldsymbol{\sigma},\boldsymbol{\sigma'}}\rho_{\boldsymbol{\sigma}, \boldsymbol{\sigma'}}(0)\ketbra{\boldsymbol{\sigma}}{\boldsymbol{\sigma'}}\ee^{-\Upsilon_{\boldsymbol{\sigma},\boldsymbol{\sigma'}}^2\Gamma(t_g)},
\end{equation}
with $\Gamma(t_g)=\frac{1}{2}\mathbb{E}[\tilde{\Delta}^2]$, and 
\begin{equation}
\tilde{\Delta}=\int_0^{\frac{t_g}{2}}\tilde{\delta}(t')\mathrm{d}t'-\int_{\frac{t_g}{2}}^{t_g}\tilde{\delta}(t')\mathrm{d}t'.
\end{equation}

Here we have assumed that the stochastic process is a zero-mean, stationary Gaussian process. By identifying the autocorrelation function $C(t-t')=\mathbb{E}[ \tilde{\delta}(t) \tilde{\delta}(t')]$, and switching to frequency domain via $C(t-t')=\int_{-\infty}^{\infty}\frac{{\rm d}\omega}{2\pi} S_\delta(\omega)\ee^{\ii\omega (t-t')}$ in terms of the power spectral density $S_\delta(\omega)$~\cite{RevModPhys.82.1155, varona2024lindbladlikequantumtomographynonmarkovian}, we can perform the time integrations. We obtain
\begin{equation}
\label{eq: gamma_parameter}
\Gamma(t_g)=\int_{-\infty}^{\infty}\mathrm{d}\omega S_\delta(\omega)F(\omega,t_g)
\end{equation}
where $F(\omega,t_g)$ is the filter function associated with the Hahn-echo sequence~\cite{Biercuk_2011, PhysRevLett.103.040501}
\begin{equation}
F(\omega,t_g)=\frac{4}{\pi\omega^2}\sin^4\left(\frac{\omega t_g}{4}\right).
\end{equation}

Including magnetic-field noise modifies the average gate fidelity by introducing, through~\eqref{eq: dephasing_state}, an additional factor multiplying each contribution to the entanglement fidelity in~\eqref{eq: fidelity_e}. The resulting entanglement fidelity is therefore
\begin{equation}
\label{eq: fidelity_e_dephasing}
F_{\mathrm{e}}(U_{\mathrm{id}}^{-1}\circ U(t))=\frac{1}{16}\sum_{\boldsymbol{\sigma,\sigma'}}\ee^{-\Upsilon_{\boldsymbol{\sigma},\boldsymbol{\sigma'}}^2\Gamma(t_g)}\Lambda_{\boldsymbol{\sigma,\sigma'}}\mathrm{tr}\{\mathcal{U}_{\boldsymbol{\sigma'}}^{\dagger}(t)\mathcal{U}_{\boldsymbol{\sigma}}(t)\rho_{\mathrm{ph}}\}.
\end{equation}

We perform simulations of the gate including dephasing, for which we choose an Ornstein–Uhlenbeck (OU) frequency noise~\cite{gardiner2004handbook,Gillespie}, which is a stationary Gaussian process fully characterized by a Lorentzian power spectral density,
\begin{equation}
\label{eq: PSD_OU}
S(\omega)=\frac{ c\tau_{\mathrm{c}}^2}{1+(\omega\tau_{\mathrm{c}})^2},
\end{equation}
where $c$ is the diffusion constant and $\tau_{\mathrm{c}}$ is the correlation time of the process. Under this noise, we can integrate~\eqref{eq: gamma_parameter} and obtain
\begin{equation}
\label{eq: gamma_parameter_OU}
\Gamma(t_g)=\frac{c\tau_{\mathrm{c}}^3}{2}\left(-3-\ee^{-\frac{t_g}{\tau_{\mathrm{c}}}}+4\ee^{-\frac{t_g}{2\tau_{\mathrm{c}}}}+\frac{t_g}{\tau_{\mathrm{c}}}\right).
\end{equation}

For a gate time much larger than the correlation time, $t_g\gg\tau_{\mathrm{c}}$, the memoryless Markovian limit is recovered. This regime is characterized by the exponential decay of the coherences at a constant rate $\Gamma(t_g)=\frac{t_g}{T_2}$ and coherence time $T_2\approx 2/c\tau_{\mathrm{c}}^2$.

\begin{figure}
\centering
\includegraphics[width=\columnwidth]{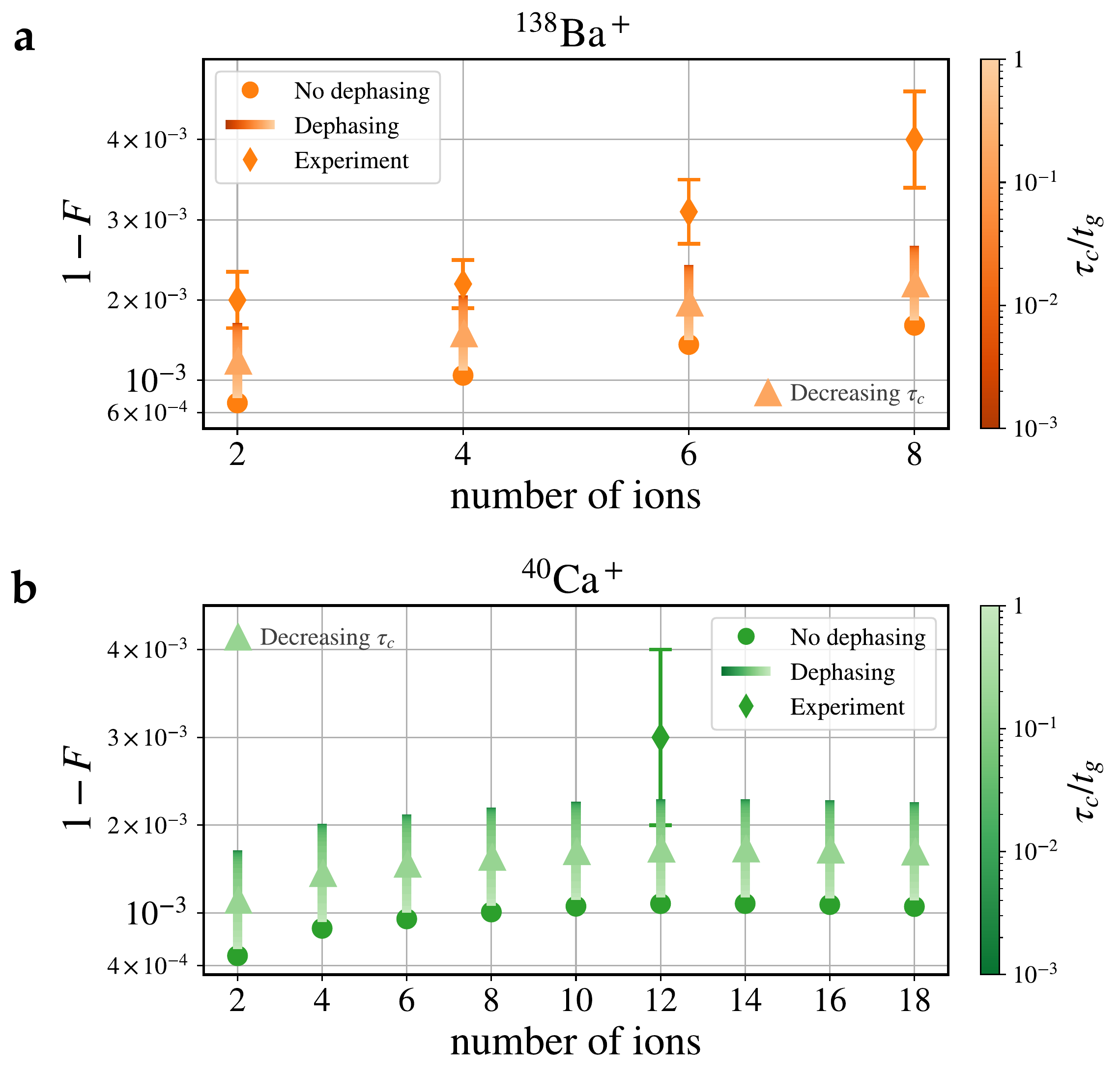}
\caption{\textbf{Comparison of the experimental state infidelity to the theoretical prediction with time-correlated  dephasing.} We evaluate expression~(\ref{eq: fidelity_e_dephasing}) analytically to obtain infidelity estimates subject to frequency fluctuations. We select correlation times ranging over the interval $\tau_c\in(10^{-3}t_g,t_g)$, adjusting the diffusion constant such that the experimental coherence time values are recovered in the Markovian limit. Such values are $T_{2,\mathrm{exp}}=0.2\mathrm{s}$ for the $^{138}\mathrm{Ba}^+$ experiment, and $T_{2,\mathrm{exp}}=0.1\mathrm{s}$ for the $^{40}\mathrm{Ca}^+$ experiment. We compute fidelity estimations for the $^{138}\mathrm{Ba}^+$ experiment up to 8 ions (\textbf{a}), with values $(\omega^{(x)},\, \omega^{(y)},\, \omega^{(z)})=2\pi\times(0.3422,\, 1.344,\, 1.304)\,\mathrm{MHz}$ for the secular trap frequencies, a detuning $\delta = 2\pi \times 20\mathrm{kHz}$ and $N_{\mathrm{loops}}=4$ loops in phase space. We do the same for the $^{40}\mathrm{Ca}^+$ experiment up to 18 ions (\textbf{b}), with values $(\omega^{(x)},\, \omega^{(y)},\, \omega^{(z)})=2\pi\times(0.480,\, 4,\, 4)\,\mathrm{MHz}$ for the secular trap frequencies, a detuning $\delta = 2\pi \times 16.7\mathrm{kHz}$ and $N_{\mathrm{loops}}=2$ in phase space. Infidelity estimates with dephasing are represented with color gradients, indicating the ratio of the correlation time relative to the gate time. The arrows indicate the direction of decreasing $\tau_{\mathrm{c}}$. Solid circular markers are employed for estimates without dephasing noise. We also plot the experimental data in~\cite{faorlin2026entanglingionsengineeredlight} with diamond markers for comparison.}
\label{fig: inf_vs_ions_dephasing}
\end{figure}

For a quantitative assessment of the effect of frequency fluctuations, we compute estimates of the state fidelity with OU frequency noise by means of~\eqref{eq: fidelity_e_dephasing} and~\eqref{eq: gamma_parameter_OU}. In Fig.~\ref{fig: inf_vs_ions_dephasing}, we plot infidelity estimates for different correlation times. For each atomic species and number of ions in the crystal, we sweep the values of $\tau_c$ over the interval $\tau_c\in(10^{-3}t_g,t_g)$. The corresponding values of the diffusion constant $c$ are obtained by assuming that the resulting coherence time matches the experimental value $T_{2,\mathrm{exp}}$ in the Markovian limit, for both the Barium experiment, $T^{\mathrm{Ba}}_{2,\mathrm{exp}}=0.2\mathrm{s}$~\cite{faorlin2026entanglingionsengineeredlight}, and the Calcium experiment, $T^{\mathrm{Ca}}_{2,\mathrm{exp}}=0.1\mathrm{s}$~\cite{Pogorelov_2021, Pal2022relaxationtimesdo}. Hence, for correlation times in the Markovian regime, such that $t_g\gg\tau_{\mathrm{c}}$, the coherences entering the infidelity in~\eqref{eq: fidelity_e_dephasing} decay exponentially, with $\Gamma(t_g)\approx\frac{t_g}{T_{2,\mathrm{exp}}}$. As the ratio $\tau_c/t_g$ increases, the non-linear dependence of~\eqref{eq: gamma_parameter_OU} on the gate time $t_g$ becomes relevant, reducing the value of $\Gamma(t_g)$ relative to its Markovian value and, consequently, the contribution of frequency fluctuations to the infidelity.

\section{Conclusions and outlook}
\label{sec. conclusions}

In this work, we have presented a thorough theoretical description of the structured-light scheme of Ref.~\citenum{faorlin2026entanglingionsengineeredlight} that demonstrated entanglement generation driving an addressed geometric LS gate through the axial vibrational modes, while using laser beams that propagate perpendicular to the ion chain. This allows minimal  crosstalk and, simultaneously, simplifies the calibration of the fully-entangling gates in 2 out of $N$ ions,  as the effective qubit-qubit interactions no longer depend on the spacing of the addressed ion pair within the chain~\cite{Leibfried2003}.

Our theoretical analysis includes closed-form expressions for the multi-mode thermal contribution to the error budget, and extensive  numerical simulations to validate  it. Our framework  corroborates the results observed in Ref.~\citenum{faorlin2026entanglingionsengineeredlight}, i.e. that high-fidelity entangling gates between pairs of ions can be realized in crystals with up to $N=12$ ions. Moreover, it predicts that the achieved errors with the same control parameters lead to similar gate fidelities without finding a severe drop in fidelity as the ion number increases to $N=20$. To do so, we have first derived a model describing the dynamics of the addressed LS gate using structured light, complementing the experimental work with a predictive theoretical analysis. We then conducted an error analysis, concluding that a co-propagating beam scheme of the addressed light gradient  is substantially less sensitive to  coupling to unwanted vibrational modes along the propagation axis. 
Our theoretical model shows that, in contrast to the counter-propagating beam scheme, a co-propagating one reduces the excitation of radial phonons through recoil which, in turn, lowers the gate sensitivity to their thermal fluctuations. To further support  these conclusions, we validated the analytical estimations with numerical simulations, including  different motional mode contributions to the error, and confirmed the  criteria  underlying the theoretical approximations made. We also incorporated the effects of dephasing noise and a Hahn-echo pulse. Our results show that thermal fluctuations and dephasing noise can account for up to roughly $90\%$ of the error budget at the operating regime of the Barium experiment in Ref.~\citenum{faorlin2026entanglingionsengineeredlight}, and about $70\%$ for the Calcium experiment. We believe that this is a remarkable result, given that we use  estimated parameters from calibration experiments but no fitting parameter.

This work contributes to the scaling of high-fidelity entangling gates in trapped-ion quantum processors, thereby helping to bring QEC logical-to-physical advantage closer to its demonstration. More particularly, we expect that the theoretical framework and analysis presented here will serve as a basis to include further estimates of other sub-leading sources of error to the budget in the future. This efforts providing useful insights for how the error budget feeds into a specific noise model and, in turn, into  QEC performance estimates that  better match future experimental implementations.

\acknowledgements
We gratefully acknowledge support by the European Union’s Horizon Europe research and innovation program under Grant Agreement Number 101114305 (“MILLENION-SGA1” EU Project), by the Austrian Science Fund (FWF Grant-DOI 10.55776/F71) (SFB BeyondC) as well as the Austrian Research Promotion Agency (FFG) under contracts No. 897481 (High-Performance integrated Quantum Computing) and No. 896213 (Ion trap arrays for quantum computing) and by IQI GmbH. We further acknowledge funding by the Intelligence Advanced Research Projects Activity (IARPA) and the Army Research Office, under the Entangled Logical Qubits programme through Cooperative Agreement Number W911NF-23-2-0216. This work has been supported by QuantERA FALCON, by FFG FO999938253, and co-funded by the European Commission. P.V., A.K. and A.B. acknowledge support from PID2021-127726NB-I00 and PID2024-161474NB-I00 (MCIU/AEI/FEDER,UE), from QUITEMAD-CM TEC-2024/COM-84, from the Grant IFT Centro de Excelencia Severo Ochoa CEX2020-001007-S funded by
MCIN/AEI/10.13039/501100011033 and under the Quantum Matter and Fields research line, and from the CSIC Research Platform on Quantum Technologies PTI-001. This work is partially funded by the European Commission–NextGenerationEU, through Momentum CSIC Programme: Develop Your Digital Talent. The authors would also like to thank Thomas Feldker, Alexander Erhard and Georg Jacob for their valuable feedback on this work.

The views and conclusions contained in this document
are those of the authors and should not be interpreted as
representing the official policies, either expressed or implied, of IARPA, the Army Research Office, or the U.S.
Government. The U.S. Government is authorized to reproduce
and distribute reprints for Government purposes
notwithstanding any copyright notation herein.

\appendix
\section{Explicit calculation of the phonon-dependent functions in the Magnus expansion}
\label{appendix 1}
In Subsec.~\ref{subsec. fidelity1} of the main text we introduced the functions $\mathcal{J}_0(t)$, $\mathcal{J}_1(t)$, $\alpha_0(t)$ and $\alpha_1(t)$. These functions result from the computation of the evolution operator~(\ref{eq: gate}) via Magnus expansion, including all contributions up to second-order in the LD parameter when resonant, and only leading order when off-resonant. These terms appear on the Magnus expansion up to fourth order. Here we include their explicit form
\begin{widetext}
\begin{equation}
\begin{split}
\mathcal{J}_0(t)=&\ii\mathcal{F}^2\left(\frac{1}{\Sigma^2}\left(\Sigma t-\sin{\Sigma t}\right)+\frac{1}{\delta^2}\left(1-10\eta^2\right)\left(\delta t-\sin{\delta t}\right)+\frac{1}{\Sigma\delta}\left(\sin{\Sigma t}+\sin{\delta t}\right)-\frac{\sin{(\Sigma-\delta)t}}{(\Sigma-\delta)}\left(\frac{1}{\Sigma}+\frac{1}{\delta}\right)\right)-\\
&-\ii\frac{5}{3}\left(\frac{\mathcal{F}}{\delta}\right)^4\eta^2\left(10\delta t+8\delta t\cos{\delta t}-16\sin{\delta t}-\sin{2\delta t}\right),\\
\mathcal{J}_1(t)=&-\ii5\left(\frac{\mathcal{F}}{\delta}\right)^2\eta^2(\delta t-\sin{\delta t}),\\
\alpha_0(t)=&\ii\mathcal{F}\left(\frac{1}{\Sigma}(\ee^{\ii\Sigma t}-1)-\frac{1}{\delta}(\ee^{\ii\delta t}-1)\right)-\ii\frac{5}{3}\left(\frac{\mathcal{F}}{\delta}\right)^3\eta^2\left(9+\ee^{-\ii\delta t}-\ee^{\ii2\delta t}+6\ii\delta t- \ee^{\ii\delta t}(9-6\ii\delta t)\right),\\
\alpha_1(t)=&-\ii20\frac{\mathcal{F}}{\delta}\eta^2(\ee^{\ii\delta t}-1).
\end{split}
\end{equation}

Equivalently, we write the functions employed in Subsec.~\ref{subsec. fidelity2}. Because we usually compute these functions at different initial times $t_0$ and with the aim of keeping a compact structure, we write them in terms of integrals stemming from the Magnus expansion, for all of which we find a primitive. The functions read

\begin{equation}
\begin{split}
\mathcal{J}(t)=&-\ii\frac{1}{2}\mathcal{F}^2\left(I_{\mathcal{J},1}(-\delta_{\mathrm{com}_x},\delta_{\mathrm{com}_x})\sum_{\substack{\alpha, m_{\alpha}\\ j}}k^{\alpha}({\gamma^{\alpha}_{m_{\alpha}, j}})^2(2{a_{m_{\alpha}}^{\alpha}}^{\dagger}a_{m_{\alpha}}^{\alpha}+1)+\frac{1}{(\gamma_{\mathrm{com}_x,0}^{x})^2}\Big(\sum_{\substack{m_x\\ j_1<j_2}}I_{\mathcal{J},2}(\delta_{m_x},\Sigma_{m_x})\gamma_{m_x,j_1}^{x}\gamma_{m_x,j_2}^{x}\right.+\\&\left.+\sum_{\substack{m_x,m_z\\ j_1<j_2}}\left(I_{\mathcal{J},3}(\delta_{m_x},\Sigma_{m_x},\nu^z_{m_z}){a^z_{m_z}}^{\dagger}+\mathrm{H.c.}\right)\gamma_{m_x,j_1}^{x}\gamma_{m_x,j_2}^{x}(\gamma_{m_z,j_1}^{z}+\gamma_{m_z,j_2}^{z})+\tan^2{\phi}(x\leftrightarrow y)\Big)\right)-\\
&-\ii\frac{10}{3}\mathcal{F}^{4}I_{\mathcal{J},4}(\delta_{\mathrm{com}_x})\sum_{j_1<j_2}((\gamma_{\mathrm{com}_x,j_1}^{x})^2+(\gamma_{\mathrm{com}_x,j_2}^{x})^2),\\
\mathcal{\gamma}(t)=&-\ii\frac{1}{2}\mathcal{F}^2\left(I_{\mathcal{J},1}(-\delta_{\mathrm{com}_x},\delta_{\mathrm{com}_x})\sum_{\substack{\alpha, m_{\alpha}\\ j}}k^{\alpha}({\gamma^{\alpha}_{m_{\alpha}, j}})^2(2{a_{m_{\alpha}}^{\alpha}}^{\dagger}a_{m_{\alpha}}^{\alpha}+1)+\frac{1}{(\gamma_{\mathrm{com}_x,0}^{x})^2}\Big(\sum_{\substack{m_x\\ j}}I_{\mathcal{J},2}(\delta_{m_x},\Sigma_{m_x})(\gamma_{m_x,j}^{x})^2\right.+\\&\left.+\sum_{\substack{m_x,m_z\\ j}}\left(I_{\mathcal{J},3}(\delta_{m_x},\Sigma_{m_x},\nu^z_{m_z}){a^z_{m_z}}^{\dagger}+\mathrm{H.c.}\right)(\gamma_{m_x,j}^{x})^2\gamma_{m_z,j}^{z}+\tan^2{\phi}(x\leftrightarrow y)\Big)\right)-\\
&-\ii\frac{2}{3}\mathcal{F}^{4}k^{x}I_{\mathcal{J},4}(\delta_{\mathrm{com}_x})\sum_{j_1<j_2}((\gamma_{\mathrm{com}_x,j_1}^{x})^2+(\gamma_{\mathrm{com}_x,j_2}^{x})^2),\\
\alpha_j(t)=&\frac{1}{2}\mathcal{F}\left((I_{\gamma,1}(\delta_{\mathrm{com}_x})\hat{\mathcal{O}}_j{a^x_{\mathrm{com}_x}}^{\dagger}-\mathrm{H.c.})+\frac{1}{{\gamma_{\mathrm{com}_x,j}^x}}\bigg(\sum_{m_x}\Big(\gamma_{m_x,j}^{x}(I_{\gamma,1}(\delta_{m_x})-I_{\gamma,1}(\Sigma_{m_x})){a^x_{m_x}}^{\dagger}-\mathrm{H.c.}\Big)\right.+\\
&+\sum_{m_x,m_z}\gamma_{m_x,j}^x\gamma_{m_z,j}^z\Big({a_{m_x}^x}^{\dagger}{a_{m_z}^z}^{\dagger}(I_{\gamma,1}(\delta_{m_x}+\nu_{m_z}^z))+I_{\gamma,1}(\Sigma_{m_x}+\nu_{m_z}^z))+{a_{m_x}^x}^{\dagger}a_{m_z}^z(I_{\gamma,1}(\delta_{m_x}-\nu_{m_z}^z))+I_{\gamma,1}(\Sigma_{m_x}-\nu_{m_z}^z))-\mathrm{H.c.}\Big)+\\
&+\tan^2{\phi}(x\leftrightarrow y)\bigg)\Bigg)+\ii\frac{2}{3}\mathcal{F}^{3}k^{x}\sum_{j_1<j_2}\left((\gamma_{\mathrm{com}_x,j_1}^{x})^2+(\gamma_{\mathrm{com}_x,j_2}^{x})^2\right)\left(I_{\gamma,2}(\delta_{\mathrm{com}_x}){a_{\mathrm{com}_x}^x}^{\dagger}+\mathrm{H.c.}\right).
\end{split}
\end{equation}

Where, in the previous expressions, we have $\gamma_{m_{\alpha},j}^{\alpha}=M^{\alpha}_{m_{\alpha},j}\eta^{\alpha}_{m_{\alpha}}$, $(k^x,k^y,k^z)=(\cos^2\phi-6,\sin^2\phi-2,1)$ and $\hat{\mathcal{O}}_j=\sum_{\alpha,m_{\alpha}}k^{\alpha}(\gamma^{\alpha}_{m_{\alpha},j})^2(2{a_{m_{\alpha}}^{\alpha}}^{\dagger}a_{m_{\alpha}}^{\alpha}+1)-k^{x}(\gamma^{x}_{\mathrm{com}_x,j})^2{a_{\mathrm{com}_{x}}^{x}}^{\dagger}a_{\mathrm{com}_{x}}^{x}$. We use the notation $(x \leftrightarrow y)$ to denote the expression inside the parentheses with every occurrence of $x$ and $y$ interchanged. Finally, we defined the integrals
\begin{equation}
\begin{split}
I_{\mathcal{J},1}(\delta_1,\delta_2)=&\int_{t_0}^t\mathrm{d}t_1\int_{t_0}^{t_1}\mathrm{d}t_2\sin(\delta_1t_1+\delta_2t_2),\\
I_{\mathcal{J},2}(\delta_1,\delta_2)=&-I_{\mathcal{J},1}(-\delta_1,\delta_2)+I_{\mathcal{J},1}(\delta_2,-\delta_1)+I_{\mathcal{J},1}(-\delta_2,\delta_2)-I_{\mathcal{J},1}(\delta_1,-\delta_1),\\
I_{\mathcal{J},3}(\delta_1,\delta_2,\nu)=&\int_{t_0}^t\mathrm{d}t_1\int_{t_0}^{t_1}\mathrm{d}t_2\left(\left(\cos(-\delta_1t_1+\delta_2t_2)-\cos(\delta_2t_1-\delta_1t_2)\right)\left(\ee^{\ii\nu t_2}+\ee^{\ii\nu t_1}\right)\right.+\\
&\left.\left(\cos(\delta_2(t_1-t_2))-\cos(\delta_1(t_1-t_2))\right)\left(\ee^{\ii\nu t_2}-\ee^{\ii\nu t_1}\right)\right),\\
I_{\mathcal{J},4}(\delta)=&\int_{t_0}^t\mathrm{d}t_1\int_{t_0}^{t_1}\mathrm{d}t_2\int_{t_0}^{t_2}\mathrm{d}t_3\int_{t_0}^{t_3}\mathrm{d}t_4\left(\cos(\delta(t_1-t_2))\sin(\delta(t_3-t_4))-\cos(\delta(t_1-t_2))\sin(\delta(t_3-t_4))+\right.\\&\left.\cos(\delta(t_1-t_2))\sin(\delta(t_3-t_4))+\cos(\delta(t_1-t_2))\sin(\delta(t_3-t_4))\right),\\
I_{\gamma,1}(\delta)=&\int_{t_0}^{t}\mathrm{d}t_1\ee^{\ii\delta t_1},\\
I_{\gamma,2}(\delta)=&\int_{t_0}^t\mathrm{d}t_1\int_{t_0}^{t_1}\mathrm{d}t_2\int_{t_0}^{t_2}\mathrm{d}t_3\left(\ee^{\ii\delta t_1}\sin(\delta(t_2-t_3))\ee^{\ii\delta t_3}\sin(\delta(t_2-t_1))\right).
\end{split}
\end{equation}

\section{Sufficiency conditions for the equality between state fidelity and entanglement fidelity.}
\label{appendix 2}
In Subsec.~\ref{subsec. fidelity1} of the main text we claimed that, for the homogeneous superposition $|\psi_0\rangle=\frac{1}{\sqrt{d}}\sum_{\boldsymbol{\sigma}}\ket{{\boldsymbol{\sigma}}}$, and when the noise channel $\tilde{\mathcal{E}}=U^{-1}\circ\mathcal{E}$ is a generalized dephasing channel~\cite{Wilde_2016}, then $F=F_{\mathrm{e}}$.

One one hand, the state fidelity reads
\begin{equation}
\label{eq: state_fidelity_appendix}
F=\braket{\psi_0\lvert \tilde{\mathcal{E}}(\ketbra{\psi_0}{\psi_0})\lvert\psi_0}=\frac{1}{d^2}\sum_{\alpha, \alpha'}\sum_{\beta, \beta'}\braket{\alpha\lvert \tilde{\mathcal{E}}(\ketbra{\beta}{\beta'})\lvert\alpha'}.
\end{equation}

On the other hand, for the entanglement fidelity $F_\mathrm{e}(\tilde{\mathcal{E}})=\bra{\Phi}(\mathbb{1}\otimes\tilde{\mathcal{E}})(\Phi)\ket{\Phi}$, with $\ket{\Phi}=\frac{1}{\sqrt{d}}\sum_{\boldsymbol{\sigma}}\ket{\boldsymbol{\sigma}}\otimes\ket{\boldsymbol{\sigma}}$, one has
\begin{equation}
\label{eq: entanglement_fidelity_appendix}
F_\mathrm{e} = \frac{1}{d^2}\sum_{\alpha, \alpha'}\sum_{\beta, \beta'}\braket{\alpha\alpha'\lvert (\mathbb{1}\otimes\tilde{\mathcal{E}})(\ketbra{\beta\beta}{\beta'\beta'})\lvert\alpha'\alpha'}=\frac{1}{d^2}\sum_{\beta, \beta'}\braket{\beta\lvert \tilde{\mathcal{E}}(\ketbra{\beta}{\beta'})\lvert\beta'}.
\end{equation}

The sufficient condition for~\eqref{eq: state_fidelity_appendix} and~\eqref{eq: entanglement_fidelity_appendix}
being equal is that the noise channel $\tilde{\mathcal{E}}$ leaves the orthonormal basis $\{\beta_i\}_{i=1}^{d}$ invariant, $\tilde{\mathcal{E}}(\ketbra{\beta}{\beta'})=c_{\beta,\beta'}\ketbra{\beta}{\beta'},$ $\forall(\beta,\beta')$. When a channel of this form satisfies the complete positivity and trace preservation constraints $c_{\beta,\beta}=1$ and $|c_{\beta,\beta'}|\leq1$, it is called a \textit{generalized dephasing channel}~\cite{Wilde_2016}.
\end{widetext}

\bibliographystyle{apsrev4-2}
\bibliography{bibliography_theo}

\end{document}